\documentclass[10pt,preprint]{aastex}
\usepackage{natbib}

\newcommand{\vrizjhk}{\protect\hbox{$V\!RIzJHK$} }

\newcommand{\be}{\begin{displaymath}}
\newcommand{\ee}{\end{displaymath}}
\newcommand{\err}[2]{\ensuremath{^{+#1}_{-#2}}}
\def\lsim{\hbox{\rlap{\raise 0.425ex\hbox{$<$}}\lower 0.65ex\hbox{$\sim$}}}
\def\gsim{\hbox{\rlap{\raise 0.425ex\hbox{$>$}}\lower 0.65ex\hbox{$\sim$}}}
\def\arcmin{\hbox{$^\prime$}}
\def\arcsec{\hbox{$^{\prime\prime}$}}
\def\fd{\hbox{$~\!\!^{\rm d}$}}
\def\fh{\hbox{$~\!\!^{\rm h}$}}
\def\fm{\hbox{$~\!\!^{\rm m}$}}
\def\fs{\hbox{$~\!\!^{\rm s}$}}

\newcommand{\nhi}{$N_{\rm HI}$}
\newcommand{\mnhi}{N_{\rm HI}}

\newcommand{\lya}{Ly$\alpha$ }

\newcommand{\kms}{km~s$^{-1}$ }
\newcommand{\cm}[1]{\, {\rm cm^{#1}}}
\newcommand{\N}[1]{{N({\rm #1})}}

\newcommand{\mkms}{{\rm \; km\;s^{-1}}}

\shorttitle{GRB 050408}
\shortauthors{Foley et al.}

\begin{document}

\title{GRB~050408: An Atypical Gamma-Ray Burst as a Probe of an Atypical
Galactic Environment}

\author{R.~J. Foley\altaffilmark{1}, D.~A. Perley\altaffilmark{1}, D.
Pooley\altaffilmark{1,2}, J.~X. Prochaska\altaffilmark{3}, J.~S.
Bloom\altaffilmark{1}, W. Li\altaffilmark{1}, B. Cobb\altaffilmark{4},
H.-W. Chen\altaffilmark{5}, G. Aldering\altaffilmark{6}, C.
Bailyn\altaffilmark{4}, C. H. Blake\altaffilmark{7}, E. E.
Falco\altaffilmark{8}, P.~J. Green\altaffilmark{7}, M.~P.
Kowalski\altaffilmark{6}, S. Perlmutter\altaffilmark{6}, K.
Roth\altaffilmark{9}, K. Volk\altaffilmark{9}}

\altaffiltext{1}{Department of Astronomy, University of California,
Berkeley, CA 94720-3411; rfoley@astro.berkeley.edu,
dperley@astro.berkeley.edu, dave@astro.berkeley.edu,
jbloom@astro.berkeley.edu, weidong@astro.berkeley.edu}
\altaffiltext{2}{Chandra Fellow}
\altaffiltext{3}{UCO/Lick Observatory, University of California, Santa
Cruz, 1156 High Street, Santa Cruz, CA 95064; xavier@ucolick.org}
\altaffiltext{4}{Department of Astronomy, Yale University, P.O. Box
208101, New Haven, CT 06520; cobb@astro.yale.edu,
bailyn@astro.yale.edu}
\altaffiltext{5}{Chicago}
\altaffiltext{6}{E. O. Lawrence Berkeley National Laboratory, 1
Cyclotron Road, Berkeley, CA 94720; galdering@lbl.gov,
mpkowalski@lbl.gov, saul@lbl.gov}
\altaffiltext{7}{Harvard-Smithsonian Center for Astrophysics, 60
Garden Street, Cambridge, MA 02138; cblake@cfa.harvard.edu,
pgreen@head.cfa.harvard.edu}
\altaffiltext{8}{F.~L. Whipple Observatory, Smithsonian Institution,
PO Box 6369, Amado, AZ 85645; falco@cfa.harvard.edu}
\altaffiltext{9}{Gemini Observatory, 670 N.~A'ohoku Place, Hilo, HI
96720; kroth@gemini.edu, kvolk@gemini.edu}

\begin{abstract} 
The bright GRB~050408 was localized by HETE-II near local midnight,
enabling an impressive ground-based followup effort as well as space-based
followup from {\it Swift}.  The {\it Swift}
data from the X-Ray Telescope (XRT) and our own optical photometry and
spectrum of the afterglow provide the cornerstone for our analysis. Under
the traditional assumption that the visible waveband was above the peak
synchrotron frequency and below the cooling frequency, the optical
photometry from 0.03 to 5.03 days show an afterglow decay corresponding
to an electron energy index of $p_{\rm lc} = 2.05 \pm 0.04$, without a
jet break as suggested by others.  A break is seen in the X-ray data
at early times (at $\sim$12600 sec after the GRB).  The spectral
slope of the optical spectrum is consistent with $p_{\rm lc}$ assuming a
host-galaxy extinction of $A_{V} = 1.18$~mag.  The optical-NIR broadband
spectrum is also consistent with $p = 2.05$, but prefers $A_{V} =
0.57$~mag.  The X-ray afterglow shows a break at $1.26 \times
10^{4}$~sec, which may be the result of a refreshed shock.  This burst
stands out in that the optical and X-ray data suggest a large \ion{H}{1}
column density of $\mnhi \approx 10^{22} \cm{-2}$; it is very likely a
damped Lyman $\alpha$ system and so the faintness of the host galaxy
($M_{V} > -18$~mag) is noteworthy. Moreover, we detect extraordinarily
strong \ion{Ti}{2} absorption lines with a column density through the GRB
host that exceeds the largest values observed for the Milky Way by an
order of magnitude.  Furthermore, the \ion{Ti}{2} equivalent width is in
the top 1\% of \ion{Mg}{2} absorption-selected QSOs.  This suggests that
the large-scale environment of GRB~050408 has significantly lower Ti
depletion than the Milky Way and a large velocity width ($\delta v >
200$\kms).
\end{abstract}


\section{Introduction}\label{s:intro}

Leading up to the launch of {\it Swift} \citep{Gehrels04}, the
astronomical community prepared for massive, multi-wavelength studies
of GRBs expected from the satellite.  Not long after the launch of {\it
Swift}, HETE-II \citep{Sakamoto05} triggered (H3711) GRB~050408 at
16:22:50.93 on 2005 April 8 ({\sc UT} dates will be used throughout this
paper).  Soon after its detection, {\it Swift} triggered a Target of
Opportunity on the GRB \citep{Wells05}.  Later, a fading optical
afterglow was detected \citep{deUgartePostigo05} and a redshift of $z
\approx 1.236$ was obtained through host galaxy emission lines and
afterglow absorption features \citep{Berger05,Prochaska05:050408}.  Radio
observations were also obtained but no transient was found
\citep{Soderberg05}. The X-ray afterglow \citep{Wells05} was observed
over several epochs with Swift, leading to an initial inference of a break
\citep{Godet05}, that was later retracted \citep{Capalbi05}. Finally with
all the XRT data, the Swift team suggested a jet break at $t_{\rm break} =
(1.2 \pm 0.5) \times 10^5$\, sec after the GRB trigger \citep{Covino05}.

We present light curves of the optical, infrared, and X-ray afterglows
in Sections~\ref{s:opt} and \ref{s:xray}.  A detailed analysis of
these afterglows is presented in Section~\ref{s:aglow}.  An analysis
of the optical and X-ray afterglow spectra is presented in
Sections~\ref{s:xray} and \ref{s:oabs}.  From the absorption in these
spectra we are able to place lower limits on the metallicity and the
hydrogen column of the host galaxy.  Throughout the paper, the
concordance cosmology of $\Omega_\lambda = 0.71$, $\Omega_m = 0.29$,
and $H_0 = 71$ km s$^{-1}$ Mpc$^{-1}$ is used. Though all measurements
reported herein are consistent with our preliminary reports in the GCN
(GRB Coordinates Network) Circulars, these measurements supersede
those in the Circulars.

\section{The Optical-Infrared Afterglow}\label{s:opt}

At approximately 18:50 on 8 April 2005, 2.4 hours after the burst,
\citet{deUgartePostigo05} detected the optical afterglow of GRB~050408.  From
this time until 20:10 on April 13, the afterglow was monitored, with
many groups reporting preliminary magnitudes and upper-limits in the
GCN
Circulars\footnotemark\footnotetext{\url{http://gcn.gsfc.nasa.gov/}}.
Here we report observations from the Keck and Magellan telescopes and
perform our own reductions of the Swift UVOT data, as described
below. All observations are summarized in Table~\ref{t:gcnphot}.

\subsection{Magellan Optical Imaging}

Our imaging with the IMACS instrument \citep{Bigelow98} on the
Magellan~I 6.5-m (Baade) Telescope began at 00:12 on 9 April 2005,
about 470 minutes after the burst.  Two 180-second exposures of the
burst field were taken with the $R_{c}$ filter and three 180-second
exposures with the $I_{c}$ filter.  Images were reduced in the
standard manner using dome flats acquired on the night of the imaging.
We performed photometry using a sample of ten reference stars
(Table~\ref{t:refstars}), six objects in the immediate vicinity of the
burst previously identified as stars by the Sloan Digital Sky Survey
(SDSS), and four additional objects in the field identified as stars
by \citet{Henden05}\footnotemark\footnotetext{\url{ftp://ftp.nofs.navy.mil/pub/outgoing/aah/grb/grb050408.dat}}
but also present in SDSS.  We use Sloan magnitudes from the SDSS
archive\footnotemark\footnotetext{\url{http://cas.sdss.org/astro/en/tools/search/}}
of all ten stars to calibrate our $R$ and $I$ instrumental magnitudes
to absolute magnitudes in $g'$, $r'$, and $i'$, and then convert back
to the Cousins system using the transform equations of
\cite{Smith02:SDSSfilters}.  Based on an astrometric comparison with
the 2MASS catalog we find the GRB occurred at position $\alpha({\rm
J2000}) =$ 12\fh02\fm17\fs.328 and $\delta({\rm J2000}) =$
+10\fd51\fm09\arcsec.47, with an error relative to the International
Coordinates Reference System (ICRS) of 250\, mas in both coordinates.

\subsection{Keck Optical Imaging}

\begin{figure}
\plotone{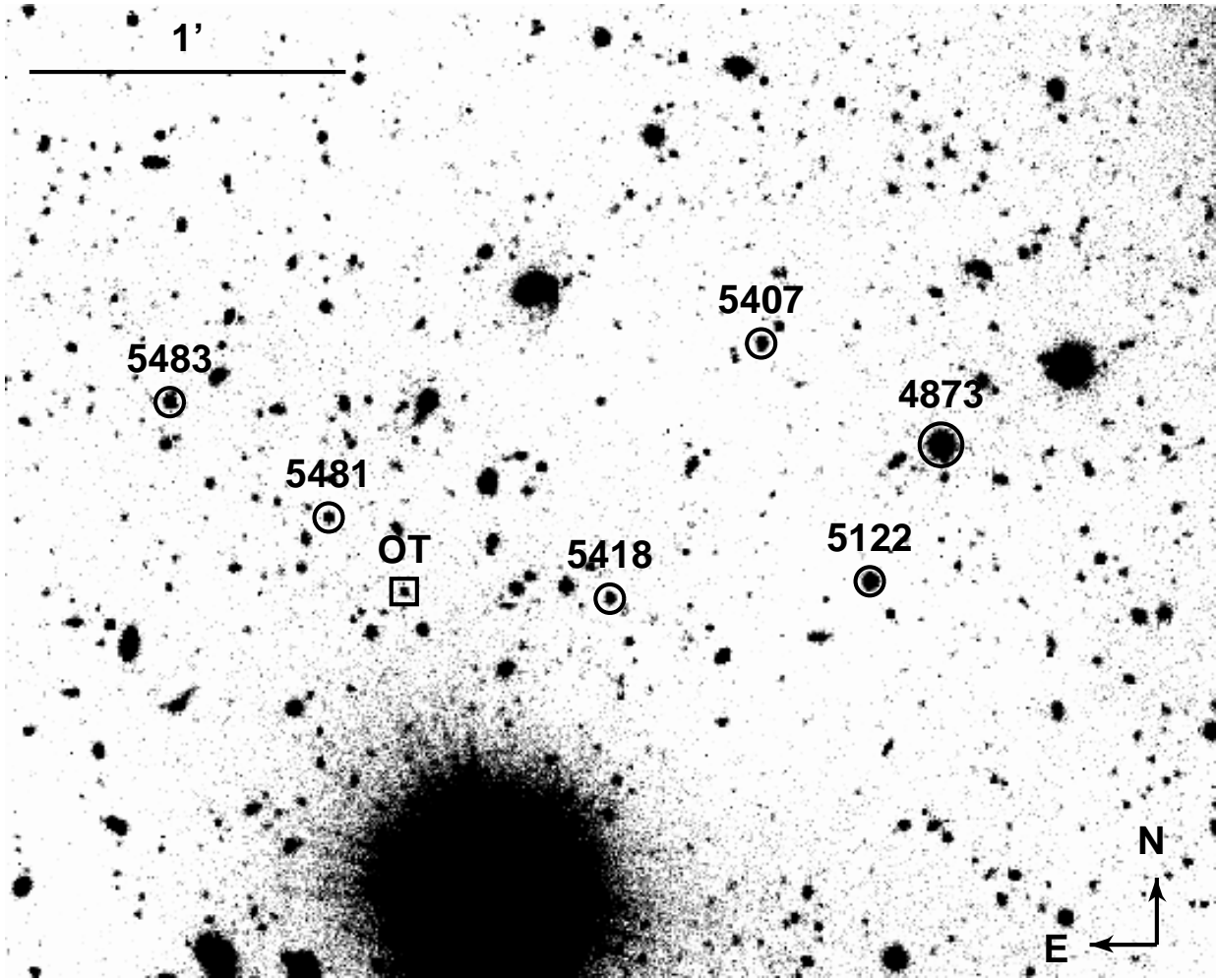}
\caption{A Keck $R$-band image of the field of GRB~050408.  The
optical transient and nearby reference stars (see
Table~\ref{t:refstars}) are marked.}\label{f:finder}
\end{figure}

Keck imaging was acquired through the UC Target of Opportunity (ToO)
program (PI Hurley) on the Keck~I 10-m telescope with the dual-beam
Low-Resolution Spectrograph Imager (LRIS; \citealt{Oke95}).  Five
60-second exposures each were taken simultaneously in $V$ and $R_c$
filters (using the D680 dichroic) beginning at 08:11 on 11 April 2005,
(2.62 days after the GRB), although because the GRB fell on a chip-gap
in one $V$-band exposure only four were used in the final analysis.
The co-added $R_{c}$-band image is shown in Figure~\ref{f:finder}.

Photometry on the Keck images was performed using the same procedure
as the Magellan images.  Because the imaged field was offset 63\arcsec
S and 5\arcsec W relative to the Magellan exposure, we use a different
sample of Henden stars, but the other six sources used for calibration
are the same.

We examined the Keck imaging for the possibility that the host galaxy
might be detected by comparing the FWHM of the afterglow with those
PSFs of field stars. There is no evidence for extension in these
images and we thus find no evidence for a host brighter than $V = 27$
mag, which corresponds to $M_{V} > -18$.

\subsection{UVOT Reductions}

The {\it Swift} Ultra-Violet/Optical Telescope (UVOT) observed the
field of GRB 050408 starting at 17:07 on 8 April 2005, 44.3 minutes
after the burst.  A series of images were obtained for the GRB field
in various filters.  An additional 11 batches of UVOT observations
were performed for GRB 050408 in the month following the GRB.

Initial results from these observations were reported by the {\it
Swift}/UVOT team \citep{Holland05}.  They reported a possible
detection ($U$ = $21.30^{+0.45}_{-0.32}$~mag) in a co-added $U$-band
image with a total exposure time of 2927 s, though they note that the
detection is marginal. They also reported no detection in other filters
and provided limiting magnitudes for the co-added images. However, the
times of the center point of the co-added images were not specified.

We retrieved the UVOT data on GRB 050408 from the {\it Swift}
Quick-look
archive\footnote{http://swift.gsfc.nasa.gov/cgi-bin/sdc/ql?} and
performed photometry using the calibration results and the photometry
recipe for {\it Swift}/UVOT from \citet{Li05:UVOT}. For the $U$-band
data, a careful inspection of all the data combined from the first two
days after the burst does not convincingly demonstrate the existence
of any object down to a limiting magnitude of $U = 21.40$~mag.  We also
re-analyzed the five $V$-band exposures from UVOT, and find no
detections to the limiting magnitudes (3$\sigma$) listed in
Table~\ref{t:uvotobs}.

\subsection{Infrared Photometry}

Infrared imaging of the field of GRB\,050408 from both the southern
(CTIO) and northern (Mt Hopkins) 2MASS 1.3m telescopes was obtained on
the first night of the burst.  The ANDICAM\footnote{http://www.astronomy.ohio-state.edu/ANDICAM  ANDICAM is
operated as part of the Small and Moderate Aperture Research Telescope
System (SMARTS) consortium. http://www.astro.yale.edu/smarts} instrument mounted on the
1.3m telescope at Cerro Tololo Inter-American Observatory (CTIO)
started observations at 03:15 on 9 April
2005.  Images
were obtained with a dual-channel camera that allows for simultaneous
optical and IR imaging.  Both optical and IR images are double-binned
in software to give an optical pixel scale of 0.27 arcsec pixel$^{-1}$
and an IR pixel scale of 0.37 arcsec pixel$^{-1}$. While standard
optical integrations are underway, the ANDICAM instrument allows IR
images to be ``dithered'' by the slight adjustment of three tilt axes
of an internal mirror.  A combination of 6 telescope re-points and 5
internal dithers were used to obtain 6 separate 360-second $I$-band
images and 30 separate 60-second $J$-band images per data set.

The Peters Automated Infrared Imaging Telescope
(PAIRITEL\footnotemark\footnotetext{\url{http://www.pairitel.org}})
started observations at 9 April 04:03:27, 11.7\,hr after the
GRB.  $J$, $H$, and $K_{s}$ band images were acquired simultaneously with
3 NICMOS3 arrays in double correlated reads with individual exposure times
of 7.8 sec.  Each image consists of a 256$\times$256
array with a plate scale of 2 arcsec pixel$^{-1}$. In a given epoch
the telescope is dithered every 3 exposures, allowing for a sky frame
appropriate for every image, derived from a star-masked median stack
of images before and after, to be created by the pipeline
software. The offsets between images are determined by a
cross-correlation and reduced images were then subsampled and stacked with a
resolution of 1 arcsec pixel$^{-1}$. The effective seeing over all the
epochs was approximately 2.3\arcsec\ FWHM.  A stack of all
offset-shifted epochs revealed a faint IR source at the location of
the optical afterglow. 

On the stacked images, we ran SExtractor\footnotemark\footnotetext{\url{http://sextractor.sourceforge.net/}} to find the instrumental
magnitudes in a 2.5 arcsec radius aperture.  These magnitudes were used
to find an absolute zeropoint uncertainty (0.02\,mag in all bands),
using more than 20 stars in common stars with the 2MASS catalog in each
band.  The transient was easily detected and well isolated in $H$ and $K_s$ but was marginally blended in the $J$.  As such, we used the average transient position
from the $H$ and $K_s$ image to determine the $x$,$y$ position in the
$J$ band image.  Fixing this center, we used IRAF/PHOT to determine the
aperture magnitude (with the 2MASS zeropoint).  For field stars, we
confirmed that IRAF/PHOT and SExtractor gave the same results within the
errors.  The $JHK_s$ magnitudes from PAIRITEL are reported in
Table~\ref{t:gcnphot}.

\subsection{Photometry from the Literature}

To produce light curves, numerous additional reported measurements
were taken from the GCN Circulars.  Optical measurements (including
upper limits) were retrieved from the circulars in all bands where
detections were reported: $B$, $V$, $R$, $I$, $J$, and ``$Z$''
(\citealt{Flasher05}, interpreted as Sloan $z'$) via
GRBlog\footnotemark\footnotetext{\url{http://grad40.as.utexas.edu/grblog.php}.
GRBlog provides a query mechanism for GCN Circulars and their
meta-data. } and the resulting table was screened for errors caused by
the automatic parsing of the circulars.  We removed duplicate reports,
as well as one observation from \citet{Milne05}, which suggested a 1
magnitude brightening in the $I$ band more than 0.5 days after the
burst (this was not seen in any other bandpass). We also culled the
$B$-band measurement from \citet{Milne05}, which was brighter in flux
than simultaneous measurements at longer wavelengths.  We replaced
data that had been superseded by later analysis for the UVOT limiting
magnitudes and Magellan observations, and added the Keck and PAIRITEL
magnitudes.

All observations used in our subsequent fitting are listed in
Table~\ref{t:gcnphot}.

\section{The X-ray Afterglow}\label{s:xray}

The {\it Swift} XRT \citep{Burrows00} began observations of GRB 050408
at 16:34 on 8 April 2005, approximately 672 sec after the HETE-II
trigger \citep{Sakamoto05}.  The XRT operates in a variety of
different observing modes, and many were used throughout the
observations.  Unfortunately, the first 1.8~ks of observations were
spent on a certain mode (the ``Low Rate Photodiode'' mode) that was
not useful for this GRB.  The ``Photon Counting'' mode observations,
which retain full imaging and spectroscopic resolution, began at
17:05:24.  As reported by \citet{Wells05}, these XRT data revealed a
fading X-ray source in the HETE-II error circle. In the ensuing weeks,
{\it Swift} observed the GRB a dozen times.  A log of the Photon
Counting mode observations is found in Table~\ref{t:xrtphot}.

We have obtained the XRT data from the {\it Swift} archive, and have
analyzed them to determine the temporal and spectral properties of the
X-ray afterglow emission.  We briefly review the data reduction, and
then we discuss the characteristics of the X-ray afterglow.

\subsection{{\it Swift} Data Reduction\label{s:xrtreduc}}

Using the Level 1 data from the {\it Swift} archive, we ran the
xrtpipeline script packaged with the HEAsoft 6.0 software supplied by
the NASA High Energy Astrophysics Science Archive Research
Center\footnote{\url{http://heasarc.gsfc.nasa.gov/}}.  We used the
default grade selection (grades 0 to 12) and screening parameters to
produce a Level 2 event file re-calibrated according to the most
current (as of 1 November 2005) calibration files in the {\it Swift}
database\footnote{\url{http://heasarc.gsfc.nasa.gov/docs/heasarc/caldb/swift/}}.
To produce images for source detection, we used the xselect software
(also part of HEAsoft 6.0), with a filter to include only counts in PI
channels 30--1000 (corresponding to photon energies of 0.3--10~keV). The PI channel to photon energy conversion was accomplished with
the redistribution file swxpc0to12\_20010101v007.rmf from the
calibration database.  The effective area of the XRT at the position
of the afterglow candidate was determined with the xrtmkarf tool,
using the correction for a point source.

\begin{figure}
\plottwo{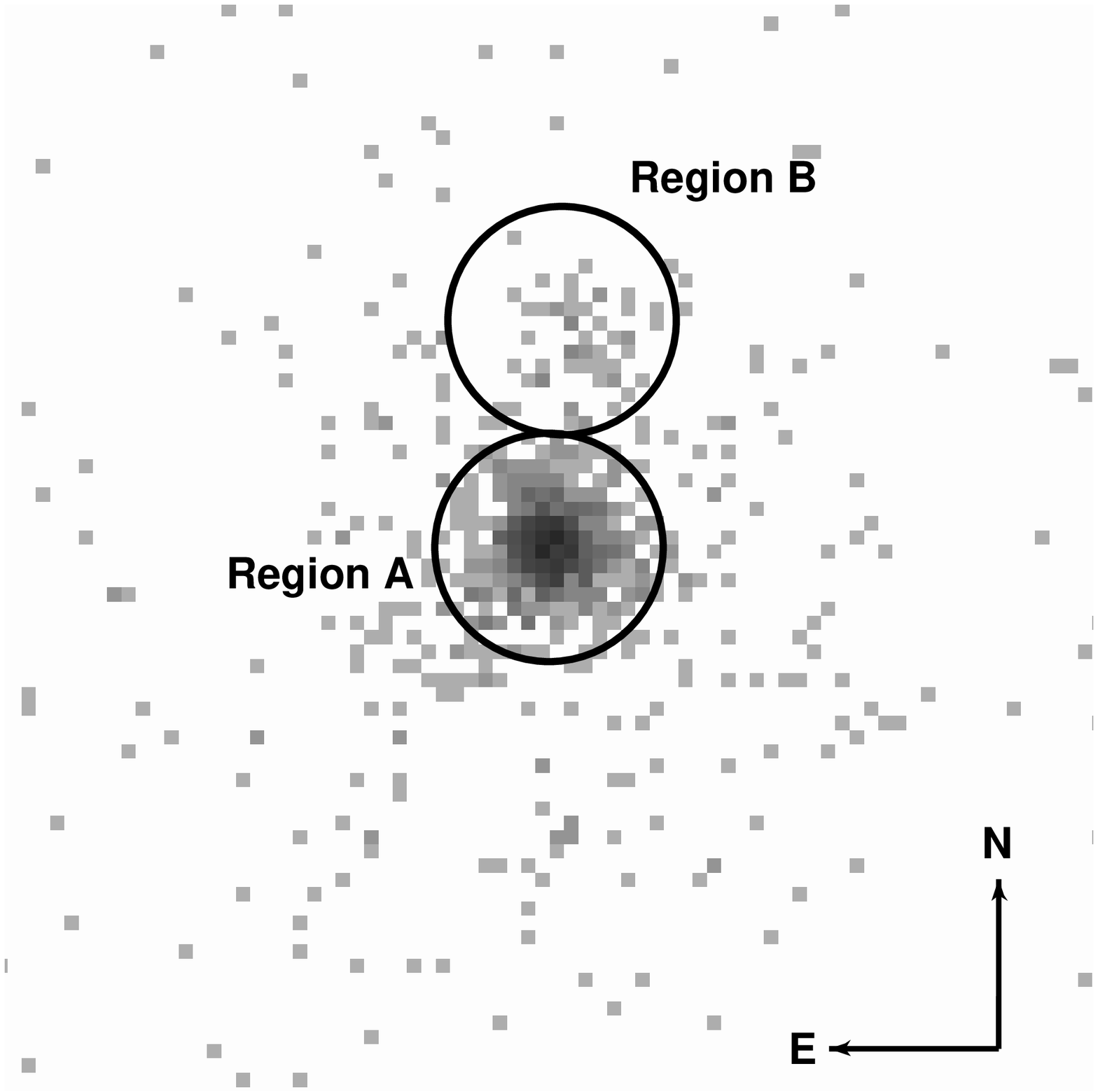}{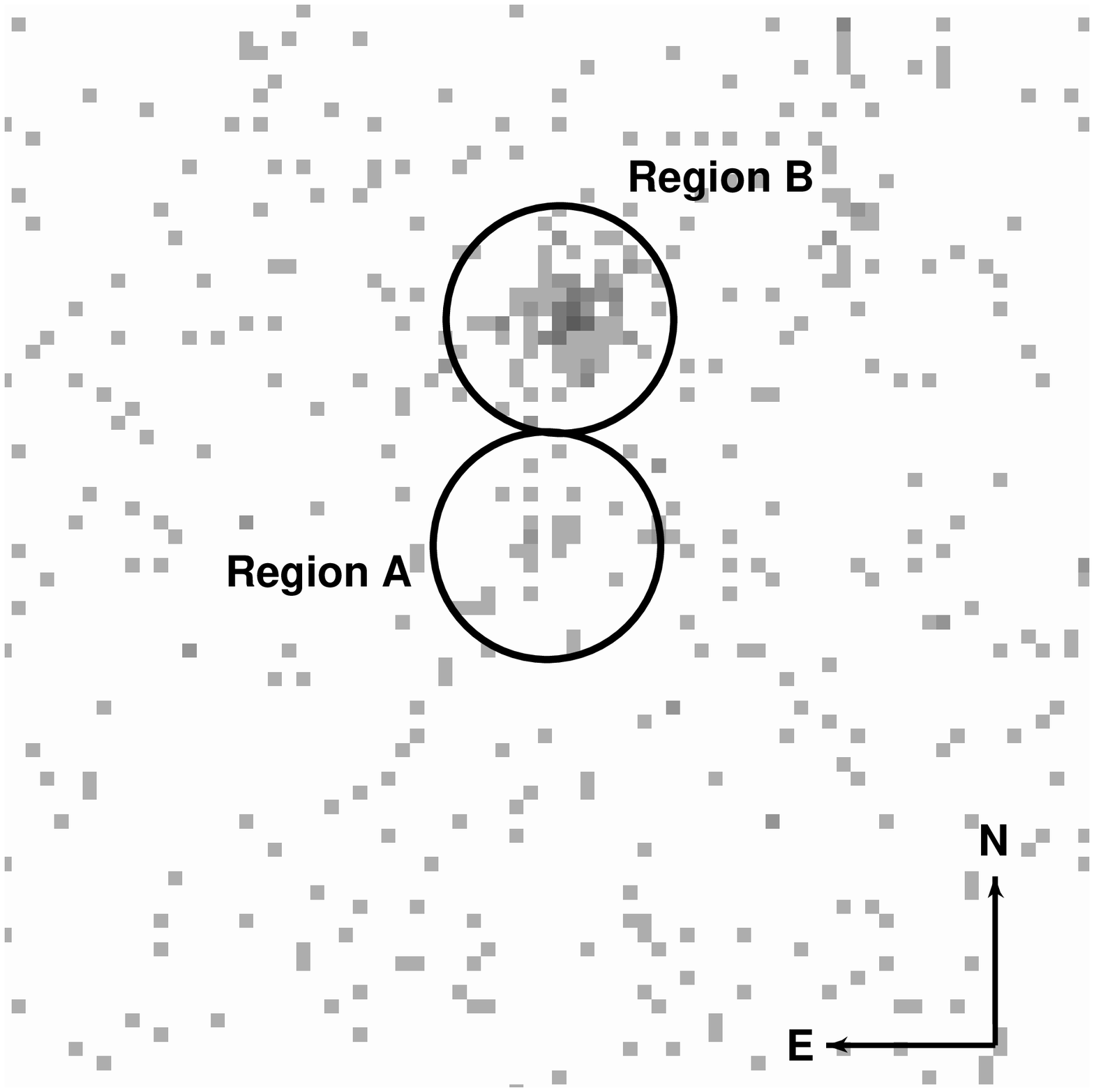}
\caption{Images of the {\it Swift} XRT data from first observation
(starting at 17:05:24 on 2005 April 8, left) and eleventh observation
(starting at 02:36:32 on 2005 May 7, right).  Each image is 3\arcmin\
on a side.  The extraction regions A (centroid of RA = 12:02:17.594,
Dec = +10:51:06.60) and B (centroid of RA = 12:02:17.448, Dec =
+10:51:44.06) are discussed in the text.}\label{f:xrtimg}
\end{figure}

Although a source extraction region of 20 pixels (47\farcs2) in radius
is recommended in the XRT Data Reduction
Guide\footnotemark\footnotetext{\url{http://heasarc.gsfc.nasa.gov/docs/swift/analysis/xrt\_swguide\_v1\_2.pdf}},
we chose a smaller extraction region (8 pixels) to mitigate the
complications due to a nearby source (designated X1) located about
37\farcs5 to the North.  Although negligible in the early
observations, this source can contribute a moderate amount of the flux
at the location of the GRB in the late
observations. Figure~\ref{f:xrtimg} shows 3\arcmin\ $\times$ 3\arcmin\
images of the XRT data from the first and eleventh observations.  We
discuss the regions A and B below.

To quantify the effects of using a smaller extraction region, we used
the XRT simulator at the ASI Science Data
Center\footnote{\url{http://www.asdc.asi.it/simulator/swift/}} to
investigate the XRT point spread function (PSF).  We simulated a very
bright X-ray source with a power-law spectrum with a photon index of
1.7 and a column density of $N_H=10^{20}$~cm$^{-2}$.  An extraction
radius of 20 pixels was found to contain $\sim$90\% of the total
counts, and an extraction radius of 8 pixels contained $\sim$70\% of
the total counts. An 8 pixel radius region located 37\farcs5 from the
source position contained $\sim$3\% of the total counts.

We can therefore represent the counts ($C$) in region A as 
\be
C_\mathrm{A} = 0.7C_\mathrm{GRB} + 0.03C_\mathrm{X1} + C_\mathrm{bkg}
\ee
where $C_\mathrm{bkg}$ is the number of expected background counts in
region A.  We use a large, source-free region to the West of the GRB
to estimate the background count density in each observation.  Using a
similar expression for the counts in region B, we can solve for the
intrinsic GRB count rate in each observation.  We list the relevant
quantities for each observation (with the first subdivided into a
number of intervals) in Table~\ref{t:xrtphot}.

\subsection{X-ray Afterglow Spectrum}

We investigate the spectral shape of the afterglow emission using a
redshifted powerlaw model with two absorption components --- a Milky
Way component set at the Galactic value of
$N_\mathrm{H}=1.81\times10^{20}$~cm$^{-2}$, and a redshifted component
local to the GRB with $N_\mathrm{H}$ allowed to vary.  We fix the
redshift of the powerlaw and the host galaxy absorption at $z=1.236$.

The spectral fitting was performed in Sherpa \citep{Freeman01} with a
hybrid Monte Carlo/Levenberg-Marquardt method.  We group the data to
have at least 15 counts per bin and fit the background-subtracted
spectra using $\chi^2$ minimization with Gehrels weighting
\citep{Gehrels86}.  To account for the contributions from the source
in region B, we fit a powerlaw to the spectrum from this source in
observation 11 (which has a relatively strong signal from region B and
a weak signal from region A).  We include this powerlaw as a fixed
component to our model of the GRB spectrum, with a normalization of
3\% of the best fit (see Section~\ref{s:xrtreduc}).

\begin{figure}
\plotone{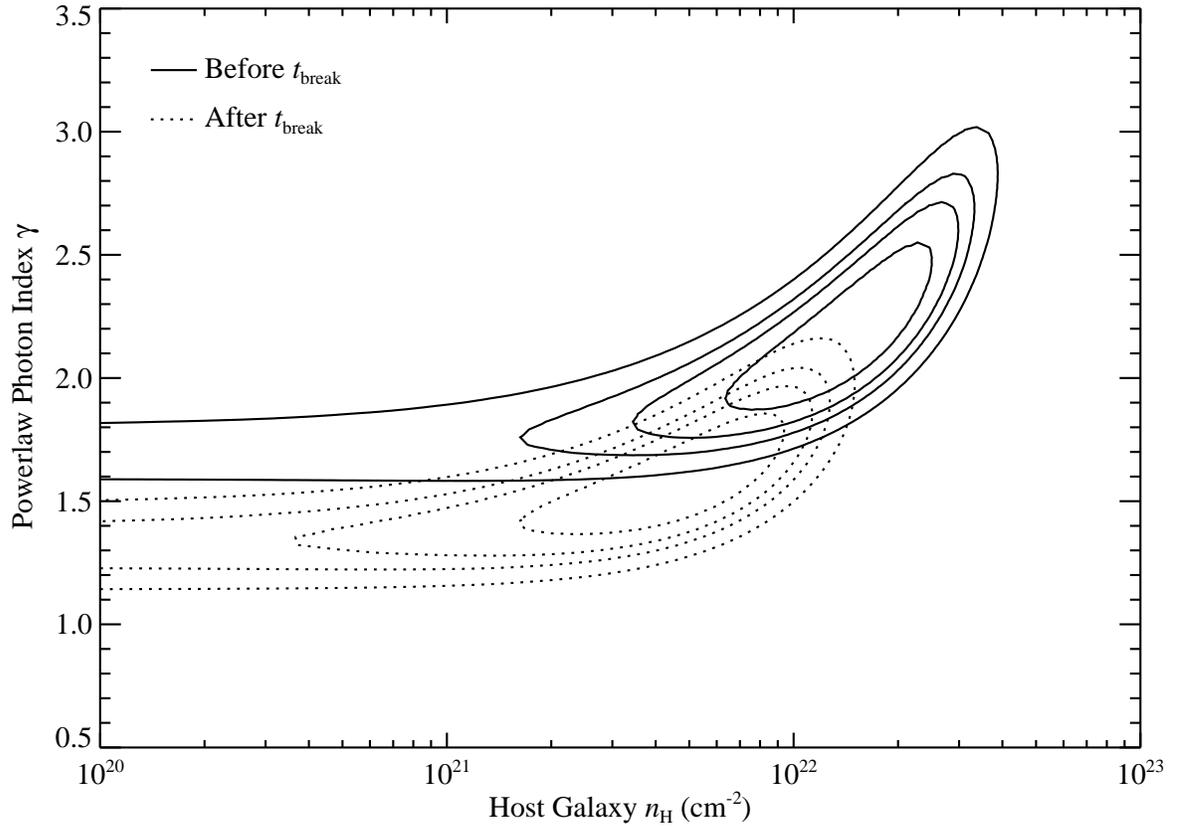}
\caption{The 68\%, 90\%, 95\%, and 99\% joint confidence intervals for
the power-law photon index of the GRB and the column density of host
galaxy.}\label{f:xrtci}
\end{figure}

We investigate separately the 0.3--10~keV data before
$t_\mathrm{break}$ (observations 1a and 1b) and after
$t_\mathrm{break}$ (observations 1c--1m, 2--12), where
$t_\mathrm{break} = 1.26 \times 10^{4}$ sec (see
Section~\ref{s:xfit}), but we find only marginal evidence for
differences in the best fit model parameters, which are listed in
Table~\ref{t:xrtspec}.  Figure~\ref{f:xrtci} plots the $\chi^2$
contours as a function of the redshifted powerlaw photon index
$\Gamma$ and redshifted $N_\mathrm{H}$ for both sets of data; contours
are drawn at 68\%, 90\%, 95\%, and 99\%.  Our best fit models had
$\chi^2$/dof of 17.2/22 (before $t_\mathrm{break}$) and 26.1/30 (after
$t_\mathrm{break}$).

As the plot shows, the host galaxy column density is poorly
constrained and is correlated with the powerlaw index.  The XRT
bandpass of 0.3--10~keV corresponds to a rest frame energy range of
0.67--22.4~keV.  Given the low statistical quality of the data, and
the fact that broadband absorption is most prominent below about 2~keV,
it is not surprising that the XRT data do not constrain the column
density very well.

\begin{figure}
\plotone{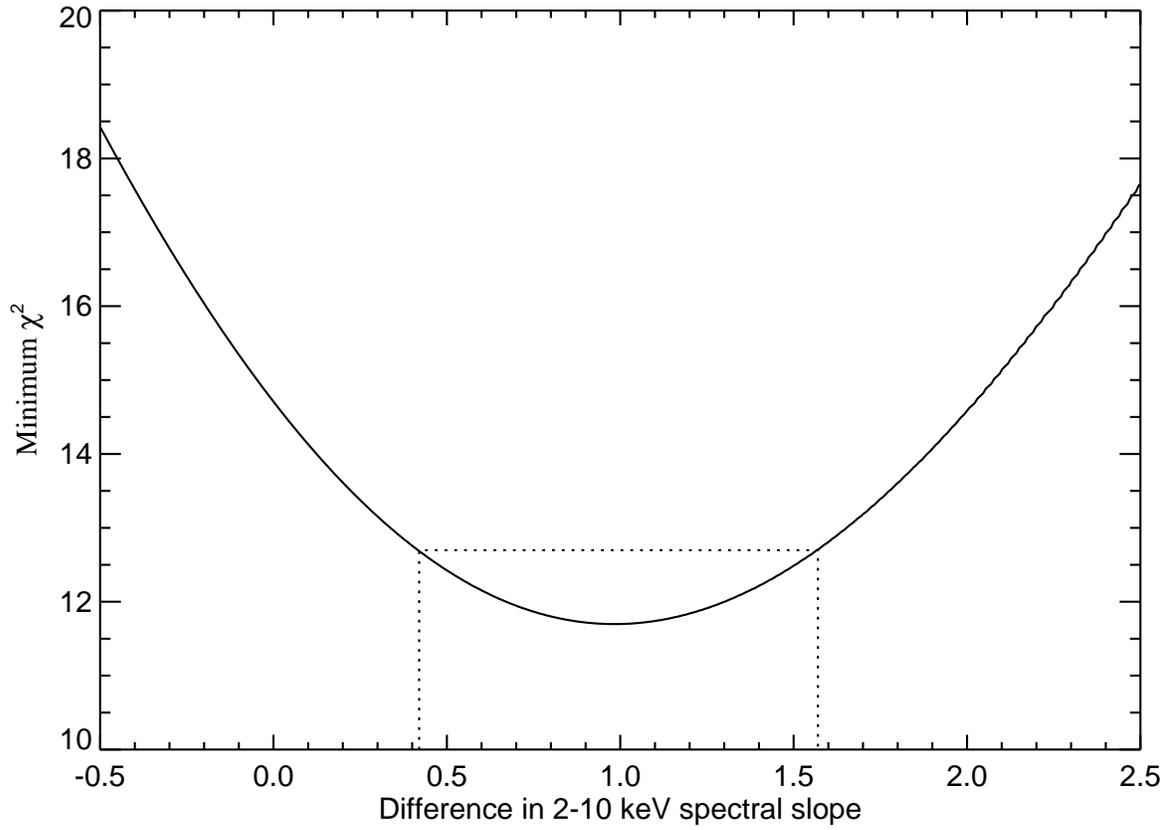}
\caption{The minimum $\chi^2$ value found in a joint spectral fit of
the 2--10 keV data before and after $t_\mathrm{break}$ as a function of
the difference in photon  indices of the two powerlaw models.  The
dotted line corresponds to the $1\sigma$ errors.\label{f:specdiff}}
\end{figure}

We investigate any intrinsic spectral differences before and after
$t_\mathrm{break}$ by considering only data above 2~keV, which is
relatively insensitive to the absorbing column to the GRB.  We perform
a 2-sided Kolmogorov-Smirnov test on the energy channel (PI)
cumulative histograms of the data before and after the break.  There
is only a 0.003 probability that they are drawn from the same parent
distribution.  To quantify this in terms of a model, we perform a
joint fit to the two data sets.  The model consists of a simple
powerlaw for each spectrum, and the free parameters are the photon
index before the break, the difference in photon index from before
until after the break, and the normalizations of the powerlaws.  We
find an acceptable fit with a $\chi^2$/dof of 11.7/23.  For these
2--10 keV data, the best fit powerlaw index before $t_\mathrm{break}$
is $2.3\pm0.5$, consistent with the best fit index found from the
entire 0.3--10 keV band (Table~\ref{t:xrtspec}).  The difference in
spectral index before and after $t_\mathrm{break}$ is found to be
$1.0\pm0.6$.  We plot the minimum $\chi^2$ found as a function of this
difference in Figure~\ref{f:specdiff}.

To estimate the X-ray flux of the GRB, we use the best fit model
before $t_\mathrm{break}$.  We need to correct the integrated model
flux since the response files generated by the XRT reduction software
assume an extraction radius of 20 pixels, in which 90\% of the flux
from a point source is contained.  Our 8 pixel radius extraction
region contains only 70\% of the flux from a point source (see
Section~\ref{s:xrtreduc}) so we multiply the best fit model flux by 1.29.
The observed, absorbed flux in the 0.3-10 keV band is
$9.0\err{4.4}{2.8}\times10^{-12}$~erg~cm$^{-2}$~s$^{-1}$.  Corrected
for Galactic absorption and absorption by the host galaxy, the
observed flux is
$15.0\err{7.4}{4.6}\times10^{-12}$~erg~cm$^{-2}$~s$^{-1}$.  This would
correspond to the emitted flux in the 0.67--22.4 keV band.  Based on
this model, the countrate to flux conversion is
$38.6\times10^{-12}$~erg~cm$^{-2}$~count$^{-1}$ (absorbed) and
$63.7\times10^{-12}$~erg~cm$^{-2}$~count$^{-1}$ (unabsorbed).

\section{Afterglow in the Context of the Synchrotron Shock
Model}\label{s:aglow}

\subsection{Optical Afterglow Fitting}\label{s:ofit}

We converted the magnitudes in Table~\ref{t:gcnphot} to absolute flux
densities using zero points from \citet{Fukugita95} and \citet{Cohen03},
assuming Cousins zero points for all $R$- and $I$-band measurements.  We
then fit these data in flux-time space with various power-law models of
the basic form
\be
F = S \left(\frac{t}{t_c}\right)^{-\alpha}.
\ee
Where $t_c$ is an arbitrary constant (defined to be 1 day) that sets
at what time the measured flux is equal to $S$.
Data without quoted errors were ignored during fitting. We ignore the
possible contribution of a constant host galaxy flux to the optical/IR
afterglow light: this appears justified by the lack of any extended
emission in the latest Keck imaging.

We fit three different models to the data: an unbroken power law where
$\alpha$ is constrained to be the same in each filter, an unbroken
power law where $\alpha$ is unconstrained, and a broken power law
where $\alpha$ and $t_{\rm break}$ are constrained to be the same in
each filter.  The data, owing largely to the steep decay implied by the
$V$-band measurements of \citet{Milne05}, formally prefer a model where
the $V$ band decay slope differs from that of the slopes in $R$ and
$I$.  However, a common power law still fits the data well:
$\chi^2$/dof = 23.6 / 24 (versus 14.7 / 21 for a variable power-law).
The broken power-law model offers negligible improvement in $\chi^2$
over an unbroken fit and is excluded to $44\%$ confidence.

\begin{figure}
\plotone{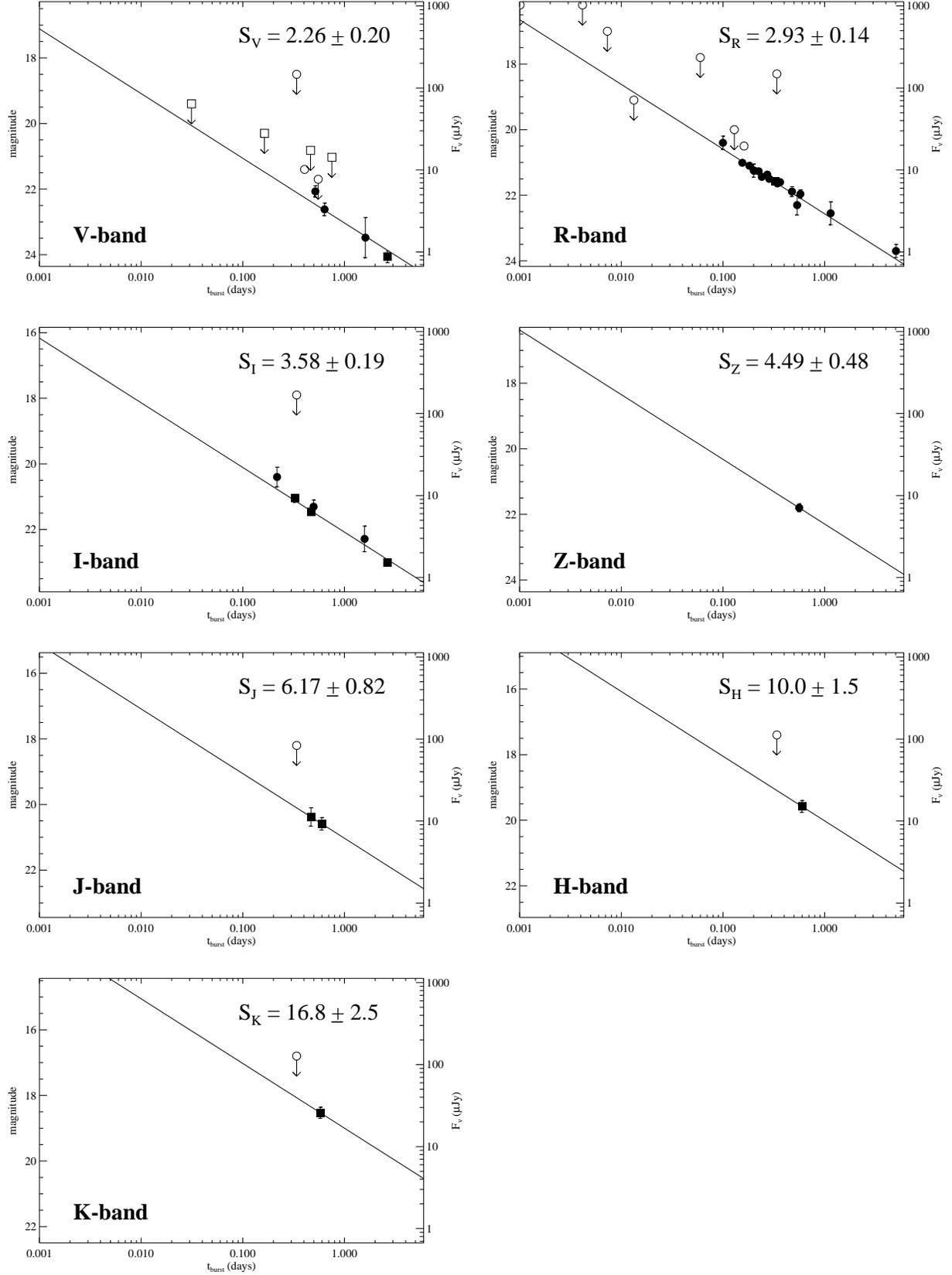}
\caption{The \vrizjhk light curves of GRB~050408.  GCN observations
are circles, our Keck, Magellan, PAIRITEL, and ANDICAM observations
presented in this paper are squares.  Filled points are detections,
while open points are upper limits.  The expected flux at $t$ = 1
day (before correction for Galactic extinction) in each band is
printed at the top right of each plot.  The best fit for the decay
constant as found in Sections~\ref{s:ofit} yields $\alpha = 0.789 \pm
0.033$.}\label{f:olc}
\end{figure}

\begin{figure}
\plotone{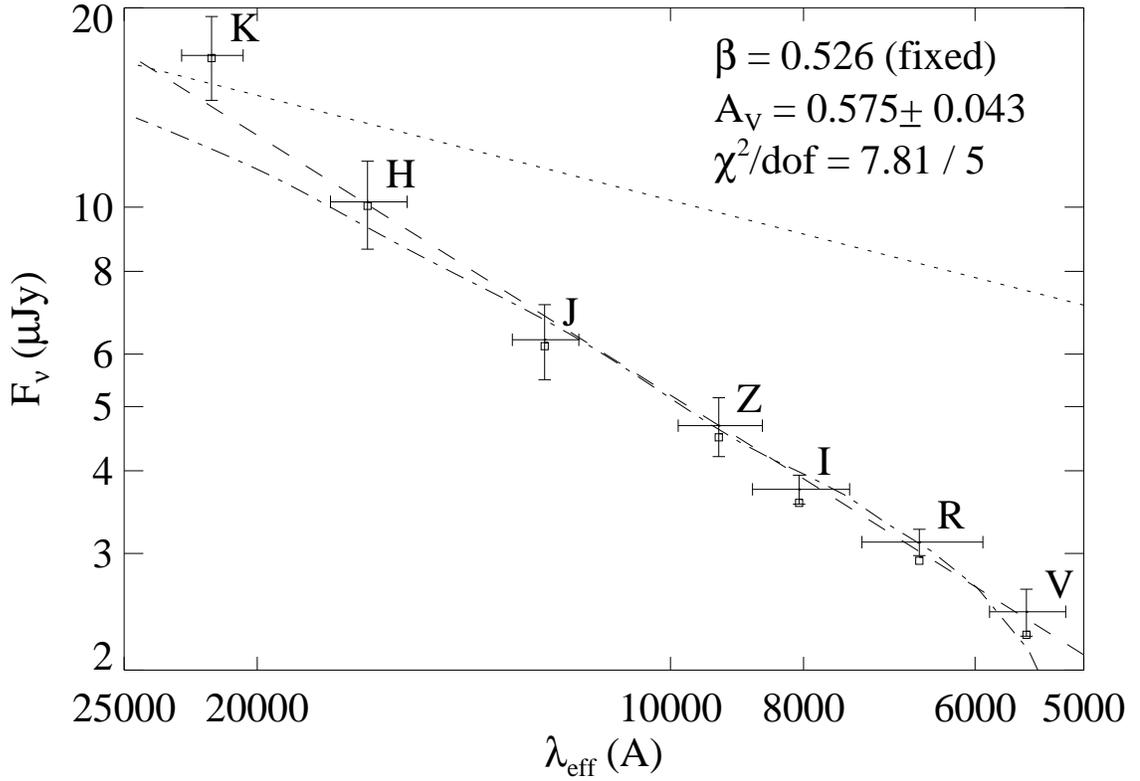}
\caption{Broadband spectrum of the afterglow of GRB050408 at one day
after the burst, assuming a uniform power-law decay in all bands.
Two model spectra are fit to the data:  an unextincted power-law
(with arbitrary spectral index $\beta$), and a power-law (with
$\beta$ = 0.53, fixed according to the value inferred from the light
curve) with extinction fit.  The unextincted fit gives a value of
$\chi^2$ = 3.19; the fit with extinction a value of $\chi^2$ =
7.81.}\label{f:broadband}
\end{figure}

If we ignore the peculiar slope of the $V$-band data (physically we do
not expect such behavior, and our own measurements do not show a
changing spectral index) and look only at the unbroken, common-alpha
model, we derive a best-fit value of $\alpha=0.789 \pm 0.033$ (see
Figure~\ref{f:olc}).  Using the linear $S$ factors and correcting for
Galactic extinction \citep{Schlegel98}, we then form a simple
broadband spectrum (Figure~\ref{f:broadband}) and fit a simple
powerlaw to this as well.  This is found to describe the data well
($\chi^2/\mathrm{dof} = 3.19/5$); the resulting value of the spectral
index, defined at $F_{\nu} \propto \nu^{-\beta}$, is then $\beta=1.30
\pm 0.10$.

Assuming a synchrotron source and certain properties of the medium in
the vicinity of the progenitor, these values for the decay constant
can be used to calculate $p$, the electron power spectrum
\citep{Sari98}.  The resulting $p$-values for a homogeneous medium are
$p_{\rm lc} = 1 + 4\alpha / 3 = 2.05 \pm 0.04$ (from the light curves)
and $p_{\rm bb} = 2 \beta_{\rm bb} + 1 = 3.61 \pm 0.19$ (from the
broadband spectrum).  These values are inconsistent, and therefore
difficult to reconcile with the simplest synchrotron model.As such, we
tried to fit a variety of afterglow models in the context of different
spectral regimes and external environments.

Following \citet{Price02}, we identify three GRB afterglow models with
different predictions about the relationship between the light curve
and spectral index:  (1) isotropic expansion into a homogeneous medium,
(2) isotropic expansion into a wind-stratified medium, and (3)
collimated expansion into a homogeneous or wind-stratified medium. The
three models share a common form for the relation between these
parameters, $\alpha + b\beta + c = 0$, where the values of $b$ and $c$
depend both on the model and whether the cooling frequency has passed
through the optical and NIR bands of our observations, for a total of
six possible closure relations.  

For each relation, we calculate the value of the closure parameter
$\alpha + b\beta = c$ and its uncertainty, as well as the predicted
electron energy spectrum index $p$, to compare the compatibility of
various models to the observations.

All models except the standard model (an isotropic medium with $\nu_c >
\nu$) predict a value of $p < 2$, a situation that is unphysical unless
there is a high-energy cutoff in the electron energy spectrum.
Furthermore, based on the calculated closure parameters, all six
relations are excluded with a confidence of at least $4\sigma$- the
spectral index predicted by these models is much less than that
observed in our optical and broadband spectra.  However, we note that
the spectral index is not corrected for host-galaxy reddening, which
the optical spectrum (discussed in Section~\ref{s:oabs})
suggest may be significant.

We examine the effect of extinction by calculating the predicted value
for the unextinguished spectral index $\beta = -(\alpha + c)/b$ for
each model and fit the observed photometric spectrum for extinction at
the host redshift, assuming a value of $R_V = 3.1$.  This provides
dramatic improvement for all models, with the best fit given by the
(equivalent) ISM-R and Wind-R models ('R' designating that the cooling
frequency is redward of the optical band; i.e.\, $\nu_c < \nu$), for
which $\chi^2/\mathrm{dof} = 5.31/5$.  For the standard assumption of
$\nu_c > \nu$, we derive a value of $\chi^2/\mathrm{dof} = 7.82 / 5$ for
a homogeneous medium or $\chi^2/{\rm dof} = 11.3 / 5$ for a
wind-stratified medium.  The jet models do not have an acceptable
$\chi^2/{\rm dof}$ whether the cooling frequency is blueward or redward
of the optical band.  The inferred value of $p$ depends only on $\alpha$
and the model, and is not extinction-dependent. As such, only the
standard homogeneous-medium model is consistent with an electron energy
spectrum that is not cut off at high energies.

Making no assumptions about the unextinguished spectral index and fitting
for the best values of $\beta$ and $A_{V}$, the optimum fit is an
unextinguished fit:  $A_{V} = 0$ and $\beta = 1.30$, for which
$\chi^2/{\rm dof} = 3.19/4$.  This is not surprising given the data:
the observed spectral slope is slightly steeper in the near-IR than in the
optical, whereas dust would be expected to steepen the spectral index in
the optical more than in the near-IR.  However, the data are not
inconsistent with the extinguished model: most of the discrepancy is due
to the single $K$-band observation. 

\begin{figure}
\plotone{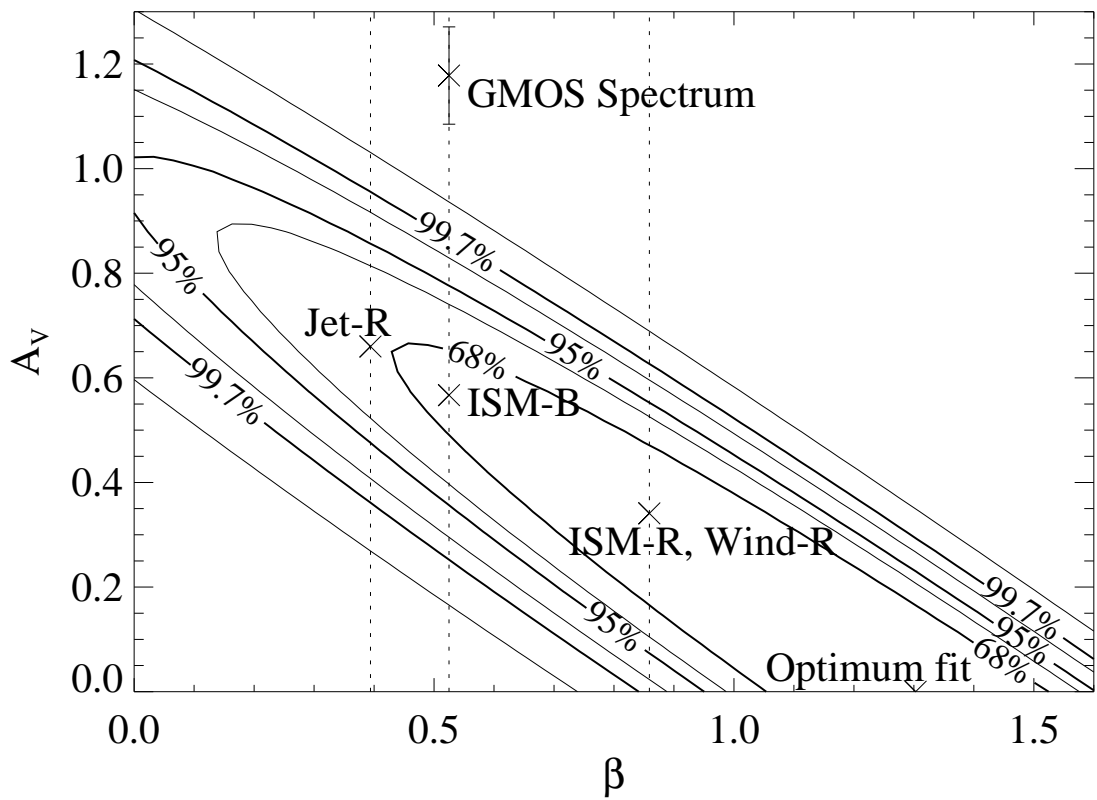}
\caption{1-, 1.5-, 2-, 2.5-, 3-, and 3.5-$\sigma$ confidence contours
for different possible values of the intrinsic $\beta$ of the
afterglow and the host-galaxy $A_{V}$, fitting to the broadband
spectrum.  In addition, we have overplotted dashed lines corresponding
to the values of $\beta$ predicted from the rate of decay of the light
curve assuming various models, and the best-fit value along those
lines.  Also overplotted is the point inferred from the GMOS spectrum
when fit with extinction.}\label{f:confcontour}
\end{figure}

Confidence contours for fits to the broadband data, varying $\beta$ and
$A_{V}$, are shown in Figure~\ref{f:confcontour}.  From this plot we see
the strong covariance between the afterglow spectral index and extinction-
and that while a steep-index, low-extinction model is preferred, smaller
values of $\beta$ are also consistent with observations if there is
sufficient extinction.  Constraining $\beta$ to the standard model of a
homogeneous medium with $\nu_{c} > \nu$, by which the afterglow is best
fit, we find $A_{V} = 0.57$~mag.  While formally the best fit yields
$A_{V} = 0$~mag, the model consistent with the afterglow decay (yielding
$A_{V} = 0.57$~mag) is within $1\sigma$ of the best fit.

\subsection{X-Ray Observations of an Early Light-Curve Break}\label{s:xfit}

\begin{figure}
\plotone{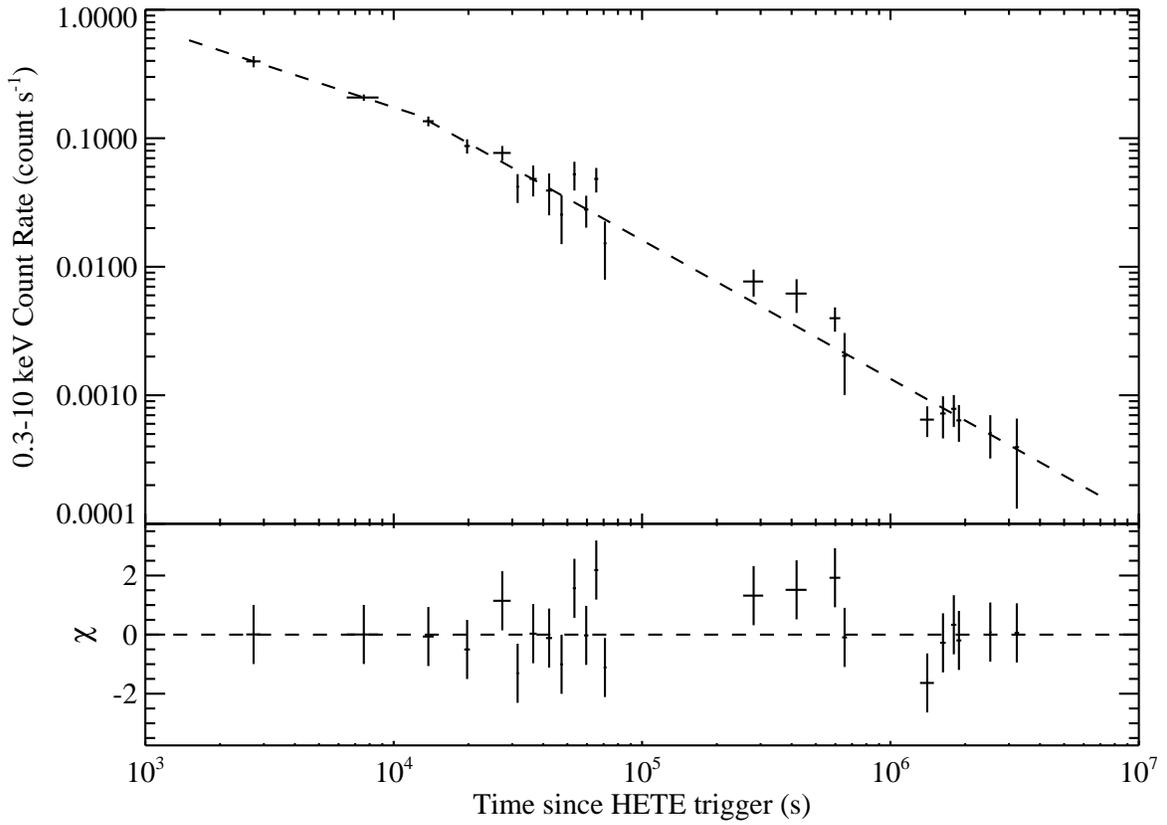}
\caption{GRB X-ray countrate (crosses) and best fitting broken power
law model (dashed line).  The horizontal bars of the crosses represent
the length of the observation interval.}\label{f:xdecline}
\end{figure}

\begin{figure}
\plotone{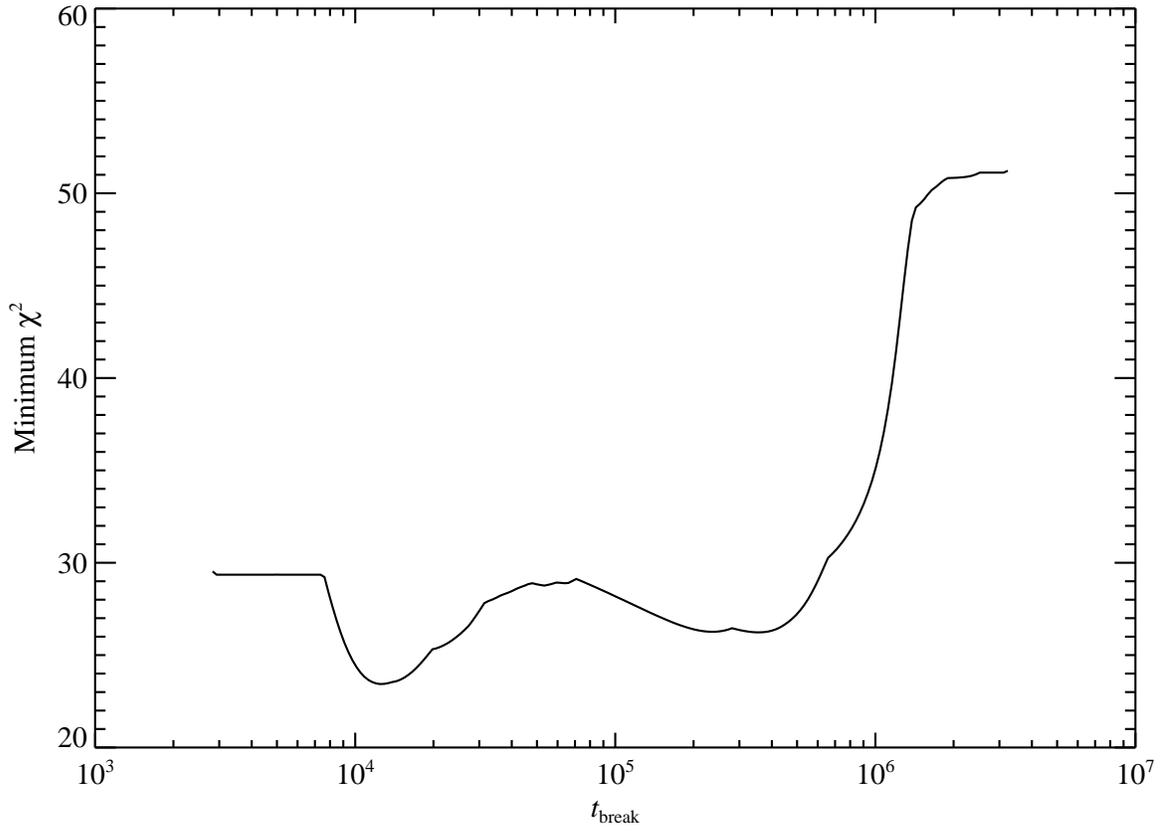}
\caption{The minimum $\chi^2$ value found as a function of
$t_\mathrm{break}$.  The flattening at early and late times is due to
the lack of points at these times, making the fit essentially a single
powerlaw.}\label{f:tbproj}
\end{figure}

We attempted to model the GRB count rate with a simple power-law
decline in time, but that gave a statistically unacceptable fit
($\chi^2$/dof = 51.2/21 corresponding to a probability of 0.025\%).  We
next tried a broken power-law model of the form
\be
R_\mathrm{x,GRB}(t) = \left\{
\begin{array}{l l}
At^{-\alpha_\mathrm{x,1}} & \quad t \le t_\mathrm{break}\\
A't^{-\alpha_\mathrm{x,2}} & \quad t > t_\mathrm{break}
\end{array} \right.
\ee
where $R_\mathrm{x,GRB}(t)$ is the GRB X-ray countrate and
$A'=A\left(t_\mathrm{break}\right)^{\alpha_\mathrm{x,2}-\alpha_\mathrm{x,1}}$.
We performed the model fitting in Sherpa, using a
hybrid Monte Carlo/Levenberg-Marquardt optimization method and a
$\chi^2$ statistic.  This hybrid method randomly samples the parameter
space 10000 times (the Monte Carlo part) for initial values and then
uses a Levenberg-Marquardt algorithm to find the local fit-statistic
minimum nearest to the starting point.  Using this method, we found an
acceptable fit ($\chi^2$/dof = 23.4/19) whose parameters are listed in
Table~\ref{t:xdecline_fit}.  The data from Table~\ref{t:xrtphot} and
the best-fit model are plotted in Figure~\ref{f:xdecline}.  We note
that the value of our break parameter $t_\mathrm{break} =
(1.26^{+1.19}_{-0.36})\times10^4$ sec is significantly lower than the
value of $(1.2\pm0.5)\times10^5$ sec found by \citet{Covino05}.  This
is most likely due to their fitting algorithm finding a local minimum
near this value. Figure~\ref{f:tbproj} plots the minimum $\chi^2$
found as a function of the parameter $t_\mathrm{break}$.

\section{Absorption Spectrum}\label{s:oabs}

\subsection{Observations and Reductions}

An optical spectrum of GRB~050408 was obtained under program
GN-2004A-Q-4, a Band 1 rapid ToO program (with carry-over status)
executed in the queue at 06:29 on 9 April 2005 April using the
Gemini~North 8-m telescope with GMOS \citep{Hook04}.  We used a 0.75
arcsecond slit, the R831 grating, OG515 order-blocking filter, and set
the central wavelength to 7330\AA.  Standard CCD processing and spectrum
extraction were accomplished with IRAF using a 1.16" aperture.  The
data were extracted using the optimal algorithm of \citet{Horne86}.
Low-order polynomial fits to calibration-lamp spectra were used to
establish the wavelength scale.  Small adjustments derived from
night-sky lines in the object frames were applied.  Using techniques
discussed in \citet{Wade88} and \citet{Matheson00}, we employed IRAF and
our own IDL routines to flux-calibrate the data and to remove telluric
lines using the well-exposed continua of the spectrophotometric standard
Feige 34 \citep{Oke90}.

\subsection{Spectral Index and Host Galaxy Dust}

Over the small wavelength range of $\sim 6280 - 8400$\AA, we were able
to fit a powerlaw spectrum to our spectrum of GRB~050408. Correcting
only for the Galactic reddening of $A_{V} = 0.081$~mag
\citep{Schlegel98}, we find $\beta_{\rm spec} = 2.11 \pm 0.29$ and
$p_{\rm spec} = 2 \beta_{\rm spec} + 1 = 5.22 \pm 0.58$.  The errors for these measurements are
statistical only.  In reality, differential light loss due to the slit
position not being at the parallactic angle \citep{Filippenko82} and
the small wavelength range will contribute additional errors.  The
$p$-value derived from the spectrum is statistically inconsistent with
that derived from both the broadband spectrum and light-curve decay
found in Section \ref{s:ofit}, $p_{\rm bb} = 3.61 \pm 0.19$ and
$p_{\rm lc} = 2.05 \pm 0.044$.  However the $p$-value derived from the
spectrum is much closer to the $p_{\rm bb}$ (within $3\sigma$) than
$p_{\rm lc}$ ($> 5\sigma$).

As discussed in Section~\ref{s:ofit}, it is possible that dust in the
host galaxy of GRB~050408 is extinguishing the spectrum, causing the
deviant spectral shape.  We performed our same fitting analysis for
the optical spectrum as above, except allowing a dust component at the
redshift of the host galaxy.  Doing this, we find that an extinction
of $A_{V} = 1.18 \pm 0.09$~mag yields a $p$-value consistent with that
found with the light curve.  The best value of $A_{V} = 1.18$~mag
yields $\beta_{\rm spec, dust} = 0.53 \pm 0.11$ and $p_{\rm spec,
dust} = 2.05 \pm 0.23$. This suggests a significant source of dust in
the host galaxy.  As shown in Section~\ref{s:ofit}, the broadband
spectrum is consistent with significant host-galaxy extinction, but
not as much as the GMOS spectrum suggests.  Again, the discrepency
between these two values are attributed to the small wavelegnth range
of the optical spectrum.

\subsection{Absorption Line Measurements}

\begin{figure}
\rotatebox{90}{
\plotone{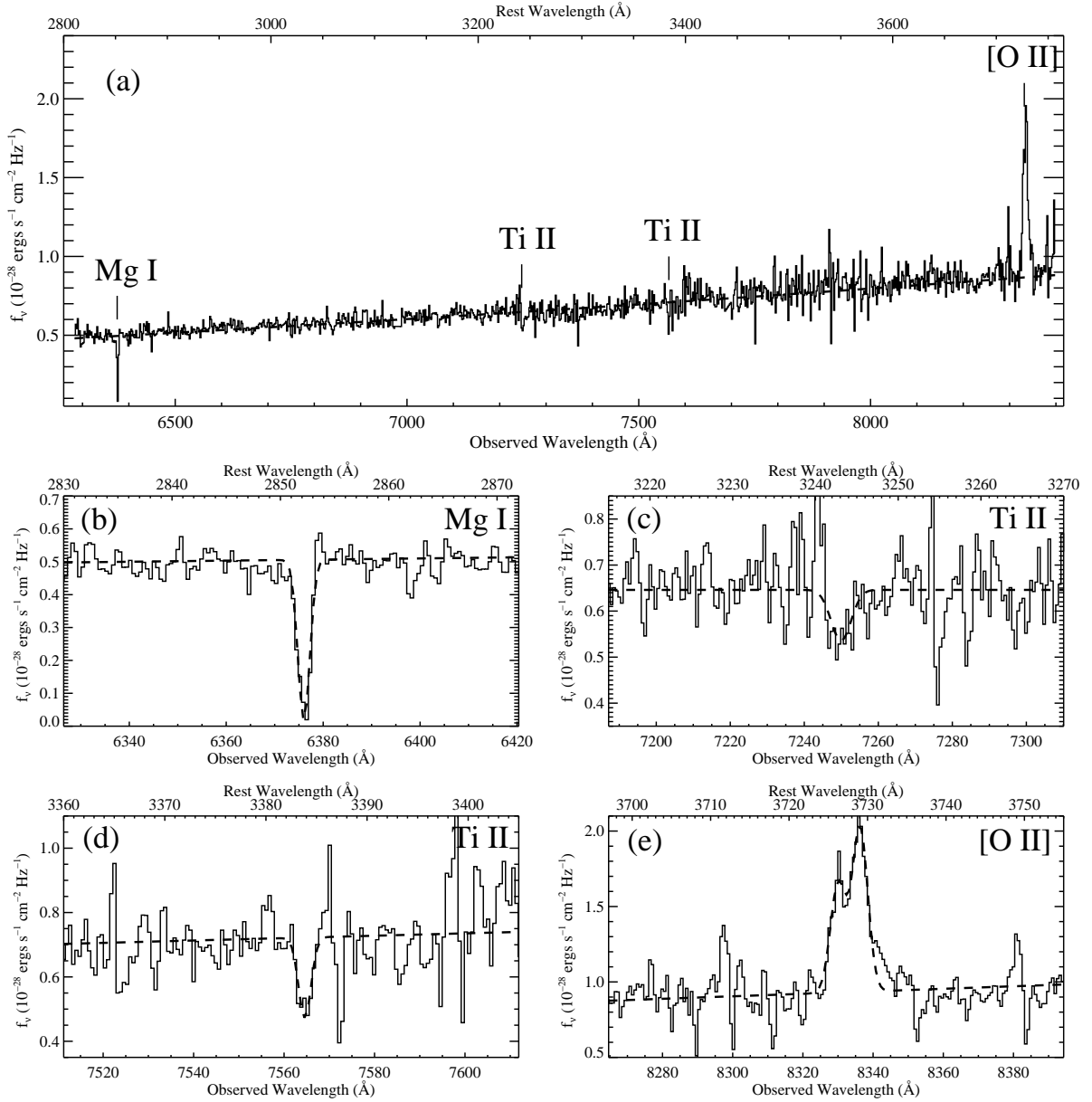}}
\caption{Optical spectrum of GRB~050408.  (a) shows the entire
spectrum with \ion{Ti}{2}, \ion{Mg}{1}, and \ion{Fe}{1} absorption
lines and [\ion{O}{2}] emission lines.  A powerlaw is fit to the
continuum and shown by the dashed line.  The powerlaw fit yields
$p_{\rm spec} = 5.22 \pm 0.58$ without host galaxy reddening and
$p_{\rm spec, dust} = 2.05 \pm 0.23$ if $A_{V} = 1.18$ in the host
galaxy.  (b)-(e) show the \ion{Mg}{1}, \ion{Ti}{2} $\lambda 3242$,
\ion{Ti}{2} $\lambda 3384$, and [\ion{O}{2}] transitions in
detail. The dashed lines in these subpanels are Gaussian fits to the
lines. The \ion{Ti}{2} $\lambda 3242$ line is blended with a bright
sky line at 7245\AA.}\label{f:spec}
\end{figure}

Figure~\ref{f:spec} presents the \ion{Ti}{2}, \ion{Mg}{1},
\ion{Fe}{1}, and [\ion{O}{2}] transitions observed in our GMOS
spectrum of the afterglow.  We have fit a local continuum at the
position of each transition and measured the rest equivalent width
$W_r$ of each feature (Table~\ref{t:ewspec}).  The errors only include
statistical uncertainty.  For the weakest transitions, uncertainty in
continuum placement will give an error comparable to the statistical
error.

Consider the \ion{Ti}{2} measurements first.  The $W_r$ values for the
four transitions are consistent with the relative oscillator strengths
and indicate the \ion{Ti}{2}~$\lambda\lambda 3242,3384$ profiles are
saturated.  We can place a conservative lower limit to the \ion{Ti}{2}
column density by adopting the $W_r$ value from  \ion{Ti}{2}~$\lambda
3384$ and ignoring corrections for line-saturation.  This gives
$N_{\rm Ti II} > 10^{13.2} \cm{-2}$.  This value is consistent with
the column density derived from \ion{Ti}{2}~$\lambda 3230$ assuming
that transition lies on the linear curve-of-growth.

Both the equivalent width and the implied ionic column density are
extraordinary.  Although Ti$^+$ is the dominant ion in neutral gas, Ti
is highly refractory \citep[e.g.][]{savage96} and only a trace amount
($\sim 1\%$) is observed in the gas-phase of the Milky Way. Therefore,
the observed equivalent widths for \ion{Ti}{2}~$\lambda 3384$ along
Milky Way sightlines are $\approx 10$m\AA\
\citep{Pettini95,Welsh97,Prochaska05}, i.e.\ 50$\times$ smaller than
that observed in this GRB afterglow. Even sightlines with $\mnhi \sim
10^{22} \cm{-2}$ have equivalent widths $W_r < 50$m\AA\
\citep{Welsh97}.  Therefore, the \ion{Ti}{2} column density along the
sightline through the GRB host galaxy exceeds  the largest values
observed for the Milky Way by an order of magnitude.  The few damped
\lya systems with \ion{Ti}{2} detections also have rest equivalent
widths $<50$m\AA\ \citep{Dessauges-Zavadsky02}.  Similarly, the LMC
has sightlines with equivalent widths $\lsim 100$m\AA\
\citep{Caulet96}.

Let us now consider the implications for the physical conditions
within the ISM of the GRB host galaxy.  The gas-phase \ion{Ti}{2}
column density that one observes is the product of three factors
(ignoring ionization corrections) -- (a) the gas column density \nhi;
(b) the metallicity of the gas [Ti/H]~$\equiv \log[\N{Ti}/\mnhi] -
\log[\N{Ti}/\mnhi]_\odot$; and (c) the depletion factor $D_{\rm Ti}
\equiv -\log[\N{Ti}_{gas}/\N{Ti}]$ -- specifically:
\be
\N{TiII} = \mnhi + {\rm [Ti/H]} - D_{\rm Ti} - 7.06
\ee
\noindent where the constant factor accounts for the solar abundance
of Ti (i.e.\ 12 - 4.94).  The first result is that the gas column
density must be large along the afterglow sightline.  Even for the
unrealistic situation that the gas has solar metallicity and is
entirely undepleted, we have $\mnhi > 10^{20.3} \cm{-2}$, i.e.\ the
sightline satisfies the \ion{H}{1} threshold which defines a damped
\lya\ system \citep[e.g.][]{Prochaska05}.  Such a large \ion{H}{1}
column density is consistent with previous measurements in GRB
afterglow spectra \citep[e.g.][]{Vreewijk03}.  The value implies the
burst originated in a gas-rich and presumably star-forming galaxy.

Second, the fact that GRB~050408 shows a much higher \ion{Ti}{2}
equivalent width than the Milky Way indicates that $\mathrm{[Ti/H]} -
D_{\rm Ti}$ is 1 dex higher in the host galaxy. Unless the GRB gas
has super-solar metallicity, the observations argue that the gas has a
significantly lower depletion level than the Milky Way ISM.
Interestingly, this result matches the conclusion for several other
afterglow spectra \citep{Savaglio04,Vreewijk03,Chen05:z=4}.  The
lower depletion of Ti could be the result of several factors. First,
the dust at high redshift may have a different composition (e.g.\ much
less Ti oxides) than the local universe.  Second, the galaxy may be
too young for the gas to have been significantly depleted from the
gas-phase.  Third, processes local to the GRB may have resulted in the
destruction of the dust grains.  These could include UV
photodissociation from OB stars in a star forming region and/or
supernova shocks or even a prompt UV flash associated with the GRB
event \citep{Waxman00}.

To this point, we have restricted the discussion to a comparison of
\ion{Ti}{2} between GRB~050408 and the Milky Way.  Therefore, one
might question whether the Milky Way has an unusual Ti depletion level
and that the characteristics of GRB~050408 are therefore not
particularly unique.   To investigate this point, we performed the following
analysis to further assess the nature of the \ion{Ti}{2} detection.
First, we compiled the set of strong ($W_r > 1.3$\AA) \ion{Mg}{2}
systems with absorption redshift $z<1.8$ identified by
\cite{Prochter05} in the Sloan Digital Sky Survey Data Release~3
quasar sample.  These quasar absorption line systems are expected to
arise in a variety of environments including galactic disks
\citep{Rao00}, galactic halos \citep{Steidel93}, and possibly galactic
superwinds \citep{Bond01}.  The key point is that the systems were
selected to have very strong metal-line \ion{Mg}{2} absorption and
also strong \ion{Fe}{2} absorption, i.e.\ metal-lines with large
$W_r$.  We then measured the $W_r$ in a 5~pixel bin centered at the
expected positions of \ion{Ti}{2}~$\lambda\lambda 3242$ and 3384.  Of
the sample of 4450\,\ion{Mg}{2} systems, only 120 showed a $3.5\sigma$
detection at the position of either \ion{Ti}{2} transition.
Furthermore, 2/3 of these `detections' were related to coincidental
absorption lines (e.g.\ \ion{Mg}{2} systems at higher redshift) or
poorly subtracted sky-lines.

\begin{figure}
\epsscale{0.7}
\rotatebox{90}{
\plotone{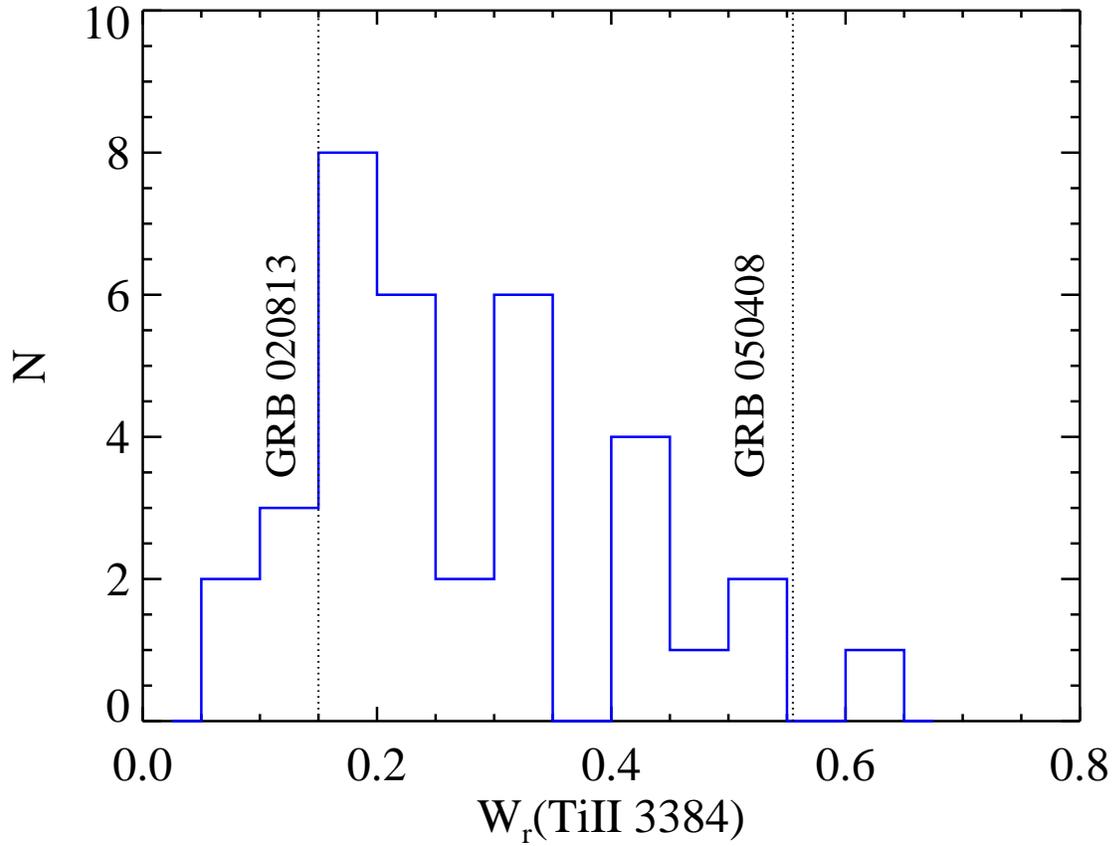}}
\caption{Histogram of $W_r$(\ion{Ti}{2} $\lambda 3384$) values
identified from 4450~strong \ion{Mg}{2} systems in the SDSS Data
Release 3 \citep{Prochter05}.  The observed values in the afterglows
of GRB~050408 and 020813 \citep{Fiore05} are given by the vertical
dashed lines.  The figure demonstrates that random sightlines through
the Universe very rarely penetrate gas with $W_r$(\ion{Ti}{2})
comparable to GRB~050408.}\label{f:tihist}
\end{figure}

Figure~\ref{f:tihist} plots a histogram of the probable detections for
\ion{Ti}{2}~$\lambda 3384$.  Our analysis indicates that fewer than
1$\%$ of the sightlines with strong \ion{Mg}{2} absorption have
correspondingly strong \ion{Ti}{2} absorption.  Furthermore, only a
handful of the positive detections have $W_r$ as large as GRB~050408
($<0.1\%$ of all strong \ion{Mg}{2} absorbers).  It is evident,
therefore, that the \ion{Ti}{2} absorption observed for GRB~050408 is
special to the GRB event.

Finally, consider the observation of \ion{Mg}{2} and the possible
detection of \ion{Fe}{1}.  Neither of these ions are dominant in
\ion{H}{1} regions because their ionization potential is less than
1~Ryd.  Therefore, it is difficult to infer physical conditions from
these features.  On the other hand, the strength of \ion{Mg}{1} is
remarkable.   Because the line is highly saturated, its equivalent
width gives a lower limit to the velocity width of the gas $\delta v >
(W_r / \lambda_r)c > 150 \mkms$.  We note that the saturated
absorption features in the afterglow spectrum of GRB~020813 also
indicate a velocity width $\delta v > 200 \mkms$ \citep[see
also][]{Fiore05}.  This appears to be a common feature of GRB
afterglow spectroscopy \citep{Vreewijk03,Ledoux05} although not a
generic feature \citep{Chen05:z=4}.  This is not, however, an expected
result in terms of the likely dynamics of the galaxy.  For example,
assume the GRB originates near the center of a rotating disk galaxy
with circular velocity $v_c$.  The maximum velocity width one would
measure is $\delta v = v_c$ and only for an edge-on sightline.  The
average value, of course, would be significantly lower.  Unless these
galaxies are relatively massive -- an assertion not supported by their
relatively low luminosities \citep{LeFloch03} -- then the observations
suggest an additional velocity field, presumably related to the GRB
environment or event.

\section{Discussion}

\subsection{Afterglow Behavior}

The synchrotron shock model \citep{Sari98} has thus far been very
successful at describing afterglow data.  The addition of breaks from
a reverse shock, cooling break, and the jet break has further
explained several afterglows.  However there are already known
inconsistencies with this model.

GRB~030329 has shown several rebrightenings in its optical light curve
which can not be explained by the synchrotron shock model.  Suggested
explanations for these rebrightenings include the refreshed shock
model \citep{Panaitescu98, Kumar00}.

The current models of GRB afterglows relying on synchrotron radiation
from a relativistic shell colliding with an external ISM does not
fully explain the afterglow of GRB~050408 without additional
considerations, such as host-galaxy extinction.  The $p$-value derived
from the afterglow decay, $p_{\rm lc} = 2.05 \pm 0.04$, is inconsistent
with the $p$-values derived from the broadband and GMOS spectra,
$p_{\rm bb} = 3.87 \pm 0.20$ and $p_{\rm spec} = 5.22 \pm 0.58$.  The
high $p$-values derived from the GMOS spectrum are determined from a
small wavelength coverage and the errors on the value are appropriate
for the wavelengths shown in the spectrum, but this region of the 
spectrum may differ
significantly from the global spectral shape.  With this
consideration, we believe the broadband and GMOS spectra have
consistent $p$-values, which are still inconsistent with the afterglow
decay $p$-value.

The most obvious physical situation that would cause a discrepancy
between the afterglow decay and the afterglow spectrum is dust (since
the decay should not be affected by this, but the spectrum will
be).  In Section~\ref{s:oabs}, we show that the absorption spectrum is
consistent with the afterglow decay if we assume a large ($A_{V} =
1.18$) host-galaxy extinction.  This value is somewhat larger than
that from the broadband spectrum ($A_{V} = 0.567 \pm 0.044$), but this
is not surprising given the discrepancy between the observed spectra.

The break observed in the X-ray afterglow is intriguing.  This change
from a shallow ($0.2 < \alpha < 0.8$) to slightly steeper ($1 < \alpha
< 1.5$) powerlaw index has been very common in {\it Swift} bursts
\citep{Nousek05}.  Although the break may be associated with a physical
mechanism associated with the afterglow (such as a minimum frequency
break), other possibilities include energy reinjection.  Our lack of
excellent sampling near the time of the break and the overall low X-ray
flux does not allow us to make further predictions.

\subsection{Hydrogen Column and \ion{Ti}{2} Abundance}

From Section~\ref{s:xrtreduc}, we found that the amount of $N_{H}$
found by fitting the X-ray spectrum is $\mnhi \approx
10^{22}$~cm$^{-2}$.  Examining the optical absorption spectrum in
Section~\ref{s:oabs}, we found $\mnhi > 10^{20.3} \cm{-2}$.  The large
equivalent widths associated with \ion{Ti}{2} in our absorption
spectrum of GRB~050408 suggest a large hydrogen column, a super-solar
metallicity, and/or a lower Ti depletion than the Milky Way.  A
super-solar metallicity seems unlikely given the redshift of the
galaxy of $z = 1.236$.  We are then forced to look at the hydrogen
column and Ti depletion.  If there is any depletion of Ti or if the
Ti/H ratio is sub-solar in the host galaxy, both of which are likely,
then a value of $\mnhi \approx 10^{22} \cm{-2}$ is quite reasonable.

The strong \ion{Ti}{2} lines are an interesting feature.  \ion{Ti}{2}
lines as strong as those in the afterglow of GRB~050408 are very rare
in \ion{Mg}{1} absorption systems ($\sim 0.1\%$).  However, these
lines are present in other GRB afterglow spectra.  This indicates that
the physical properties of the environment of these GRBs that create
the strong lines are linked to the GRB-progenitor formation or the
GRB-progenitor affected environment.  It appears that a low Ti
depletion is somehow linked to the formation of massive stars, the
environment created around such massive stars, or perhaps the event
itself.

\subsection{Line Velocities} 

The large velocities ($v \approx 150 \mkms$) implied by the absorption
lines in the spectrum of GRB~050408 are not easily explained by the
kinematics of the host galaxy.  Although it is possible that the host
is a very massive galaxy (although unlikely considering the luminosity
of $M_{V} > -18$), another scenario is that the velocity is local to
the GRB.  The presumed progenitors to long-duration GRBs, Wolf-Rayet
stars, are known to have large winds, and therefore, large line
velocities associated with them \citep{Gull05}. \citet{Mirabal03} saw
distinct systems of lines offset by $\sim 450$, $\sim 1000$, and
$> 1000$ \kms from the host redshift, which they interpreted as a
shell nebula from a Wolf-Rayet progenitor surrounding the GRB.  The
resolution of our spectrum is too low to see distinct components
within our lines, however, we can safely say that there is no strong,
distinct component at $\sim 3000$ \kms relative to the host galaxy.

\section{Conclusions}

GRB~050408 is a particularly interesting object showing both the
consistency of predicted models and showing new and extreme cases of
physical phenomena.  In particular, we have shown:

\begin{itemize}

\item The synchrotron electrons had energy index of $p \approx 2$, the
lower limit of physically acceptable systems \citep{Meszaros97,Sari98}.
This is supported directly by the optical-NIR afterglow decay and the
X-ray spectrum.  There is also indirect support (assuming particular
models) from the optical spectrum, the optical-NIR broadband spectrum,
and the X-ray afterglow decay.

\item The X-ray afterglow shows a break at $1.26 \times 10^{4}$ sec
after the burst.  This break is not attributed to a jet break.  One
possible explanation is continued energy injection.


\item The hydrogen column is very large ($\mnhi \approx
10^{22}$~cm$^{-2}$).  The optical spectrum also showed one of the
most extreme Ti-absorption systems observed.  The combination of
these facts suggest that there is an incredibly low amount of Ti
depletion in the environment of GRB~050408.  This has been noted for
other GRBs, suggesting that low Ti depletion is linked to GRB
environments, possibly due to high-mass star formation, the
environments of newly formed supernova and GRB remnants, or dust
destruction from the GRB.

\item The large velocities associated with the absorption lines are
not easily explained by the kinematics of the host galaxy.  For a
systemic velocity of $v \approx 150$ \kms, a large mass (and
possibly a special geometry) is needed.  However, we have shown
that the host of GRB~050408 is faint $M_{V} > -18$, comparable to
the LMC.  This suggests that the velocities originate close to the
progenitor, either from a wind from the Wolf-Rayet progenitor star
or older supernova explosions close to the progenitor.

\end{itemize}

\begin{acknowledgments} 

We are grateful to the Gemini observing staff. Based on observations
obtained at the Gemini Observatory, which is operated by the
Association of Universities for Research in Astronomy, Inc., under a
cooperative agreement with the NSF on behalf of the Gemini
partnership: the National Science Foundation (United States), the
Particle Physics and Astronomy Research Council (United Kingdom), the
National Research Council (Canada), CONICYT (Chile), the Australian
Research Council (Australia), CNPq (Brazil) and CONICET (Argentina)
JSB, JXP, and H-WC are partially supported by a grant from the NASA
Swift Guest Investigator Program. DP gratefully acknowledges support
provided by NASA through Chandra Postdoctoral Fellowship grant number
PF4-50035 awarded by the Chandra X-ray Center, which is operated by
the Smithsonian Astrophysical Observatory for NASA under contract
NAS8-03060.  GA, MK, and SP are supported in part by the United States
Department of Energy under contract DE-AC03-76SF000098.  The Peters
Automated Infrared Imaging Telescope (PAIRITEL) is operated by the
Smithsonian Astrophysical Observatory (SAO) and was made possible by a
grant from the Harvard University Milton Fund, the camera loan from
the University of Virginia, and the continued support of the SAO and
UC Berkeley.  RJF would like to thank Thomas Matheson for his
assistance with the GMOS data reduction.  RJF would also like to
especially thank Steve Dawson for showing examples of complicated
figures and for being a great example of a successful and
universely-admired astronomer.  The authors wish to extend special
thanks to those of Hawaiian ancestry on whose sacred mountain we are
privileged to be guests. Without their generous hospitality, many of
the observations presented herein would not have been possible.

\end{acknowledgments}

\bibliographystyle{apj}
\bibliography{astro_refs}

\begin{deluxetable}{lllllll}
\tablewidth{0in}
\tablecaption{SDSS Reference Stars\label{t:refstars}}
\tablehead{
\colhead{SDSS Obsid} &
\colhead{RA} &
\colhead{Dec}  &
\colhead{$g'$ (Mag)} &
\colhead{$r'$ (Mag)} &
\colhead{$i'$ (Mag)} &
\colhead{Images applied to}}
\tabletypesize{\small}
\startdata
587734893827063893  &  180.50328 & 10.804996 & $18.935 \pm 0.010$ & $17.583 \pm 0.006$ & $16.952 \pm 0.005$ & M\tablenotemark{a}\\
587732772665294967  &  180.53813 & 10.823279 & $18.827 \pm 0.008$ & $18.370 \pm 0.007$ & $18.208 \pm 0.008$ & M\\
587732772665294873  &  180.54322 & 10.860261 & $18.944 \pm 0.009$ & $18.601 \pm 0.008$ & $18.474 \pm 0.010$ & M, K\tablenotemark{b}\\
587734893827129348  &  180.55899 & 10.783250 & $17.741 \pm 0.005$ & $17.412 \pm 0.005$ & $17.280 \pm 0.006$ & M\\
587732772665294997  & 180.59562 & 10.899296 & $18.796 \pm 0.008$ & $18.484 \pm 0.008$ & $18.377 \pm 0.009$ & K\\
587732772665295483  &  180.58477 & 10.862593 & $22.557 \pm 0.117$ & $21.092 \pm 0.048$ & $20.152 \pm 0.031$ & M, K\\
587732772665295451  &  180.57622 & 10.856447 & $23.182 \pm 0.202$ & $22.027 \pm 0.107$ & $21.520 \pm 0.098$ & M, K\\
587732772665295418  & 180.56102 & 10.852202 & $23.003 \pm 0.188$ & $21.689 \pm 0.086$ & $21.076 \pm 0.072$ & M, K\\
587732772665295407  &  180.55286 & 10.865665 & $22.517 \pm 0.113$ & $22.271 \pm 0.129$ & $22.177 \pm 0.171$ & M, K\\
587732772665295122  &  180.54705 & 10.853076 & $20.925 \pm 0.032$ & $20.707 \pm 0.036$ & $20.534 \pm 0.043$ & M, K\\
587732772665294873  & 180.54322 & 10.860261 & $18.944 \pm 0.009$ & $18.601 \pm 0.008$ & $18.474 \pm 0.010$ & M, K\\
\hline
\enddata
\tablenotetext{a}{Magellan observations}
\tablenotetext{b}{Keck observations}
\end{deluxetable}

\begin{deluxetable}{lllll}
\tablewidth{0pc}
\tablecaption{UVOT $V$-band Observations of GRB050408\label{t:uvotobs}}
\tablehead{
\colhead{$t_{\rm start}$ (s after burst)} &
\colhead{Exp. (s)} &
\colhead{Total Exp. (s)} &
\colhead{$t_{\rm center}$ (s)} &
\colhead{Limiting Mag (3$\sigma$)}}
\startdata
2657.1   &   99.77  &    99.77s &     2707.0  &  $19.40 \pm 0.35$ \\
13803.1  &  689.35  &   689.35s &     14147.8 &  $20.30 \pm 0.25$ \\
37264.1  &  380.27  & (combined)&  \nodata    & \nodata \\
40919.1  &  899.76  & 1280.03   &     40130.1 &  $20.82 \pm 0.25$  \\
59189.1  & 897.35   & (combined)&  \nodata    & \nodata \\
70765.1  & 899.77   &  1797.12  &    64979.0  &  $21.03 \pm 0.25$ \\
\enddata
\end{deluxetable}

\begin{deluxetable}{llllll}
\tablecolumns{5}
\tablewidth{0pc}
\tablecaption{Optical/IR Observations of GRB050408\label{t:gcnphot}}
\tablehead{
\colhead{Filter} &
\colhead{$t_{\rm burst}$ (d)} &
\colhead{Mag}  &
\colhead{Ref} &
\colhead{Comments}}
\startdata
V &
0.03133 &$   19.4 \tablenotemark{a} $&    This paper & UVOT re-reduction \\
V & 0.16362 &$   20.3 \tablenotemark{a} $&    This paper & UVOT re-reduction \\
V & 0.33899 &$   18.5 \tablenotemark{a} $&   \citet{Melandri05} & \\
V & 0.40548 &$  21.4                    $&   \citet{Bayliss05} & \\
V & 0.46447 &$  20.82 \tablenotemark{a} $&    This paper & UVOT re-reduction\\
V & 0.51747 &$22.069 \pm 0.171          $&   \citet{Milne05} & \\
V & 0.55185 &$   21.7 \tablenotemark{a} $&   \citet{Bayliss05} & \\
V & 0.63747 &$22.618 \pm 0.191          $&   \citet{Milne05} & \\
V & 0.75297 &$ 21.03 \tablenotemark{a}  $&    This paper & UVOT re-reduction\\
V & 1.60747 &$ 23.48 \pm 0.612          $&   \citet{Milne05} & \\
V & 2.65673 &$24.067 \pm 0.176          $&    This paper & Keck\\
\hline
R   & 0.00007 &$ 11   \tablenotemark{a}   $&   \citet{Tamagawa05} & \\
R   & 0.00123 &$ 10.9 \tablenotemark{a}   $&   \citet{Tamagawa05} & \\
R  & 0.00101 &$ 16.2 \tablenotemark{a}   $&   \citet{Torii05} & \\
R  & 0.00416 &$ 16.2 \tablenotemark{a}   $&   \citet{Torii05} & \\
R  & 0.00730 &$    17                    $&   \citet{Torii05} & \\
R & 0.01330 &$  19.1 \tablenotemark{a}  $&   \citet{Mizuno05} & \\
R  & 0.05983 &$  17.8 \tablenotemark{a}  $&   \citet{Kuroda05} & \\
R   & 0.10000 &$  20.4 \pm 0.2            $&   \citet{Misra05} & \\
R & 0.12927 &$    20 \tablenotemark{a}  $&   \citet{Klose05} & \\
R  & 0.15500 &$ 21.01 \pm 0.07           $&   \citet{Bikmaev05} & \\
R & 0.16122 &$  20.5                    $&   \citet{deUgartePostigo05}&\\
R  & 0.18208 &$  21.1 \pm 0.05           $&  \citet{Bikmaev05} & \\
R & 0.20083 &$ 21.25 \pm  0.2           $&   \citet{Wiersema05} & \\
R  & 0.20375 &$ 21.25 \pm 0.05           $&   \citet{Bikmaev05} & \\
R  & 0.22458 &$ 21.27 \pm 0.05           $&   \citet{Bikmaev05} & \\
R  & 0.24125 &$ 21.44 \pm 0.06           $&   \citet{Bikmaev05} & \\
R  & 0.27292 &$ 21.37 \pm 0.06           $&   \citet{Bikmaev05} & \\
R  & 0.28542 &$  21.5 \pm 0.06           $&   \citet{Bikmaev05} & \\
R & 0.32580 &$21.584 \pm 0.104          $&   This paper & Magellan \\
R & 0.33899 &$ 18.3  \tablenotemark{a}  $&   \citet{Melandri05} & \\
R  & 0.34333 &$ 21.64 \pm 0.07           $&   \citet{Bikmaev05} & \\
R  & 0.36458 &$  21.6 \pm 0.07           $&   \citet{Bikmaev05} & \\
R & 0.47747 &$21.888 \pm 0.15           $&   \citet{Milne05} & \\
R & 0.53622 &$  22.3 \pm 0.3            $&   \citet{Curran05} & \\
R & 0.57747 &$21.963 \pm 0.129          $&   \citet{Milne05} & \\
R & 1.14546 &$ 22.55 \pm 0.35           $&   \citet{Kahharov05} & \\
R  & 5.03250 &$  23.7 \pm 0.2            $&   \citet{Bikmaev05} & \\
\hline
I & 0.21750 &$  20.4 \pm 0.3            $&  \citet{Curran05} & \\
I & 0.32580 &$21.048 \pm 0.123          $&   This paper & Magellan \\
I & 0.33899 &$  17.9 \tablenotemark{a}  $&   \citet{Melandri05} & \\
I & 0.46930 &$ 21.47 \pm 0.11           $&   This paper & CTIO 1.3m\\
I & 0.49747 &$21.305 \pm 0.203          $&   \citet{Milne05} & \\
I & 1.57747 &$22.288 \pm 0.39           $&   \citet{Milne05} & \\
I & 2.65673 &$23.012 \pm 0.109          $&   This paper & Keck \\
\hline
Z\tablenotemark{b} & 0.56190 &$ 21.8  \pm 0.12          $&   \citet{Flasher05} & \\
\hline
J & 0.33899 &$ 18.2 \tablenotemark{a}   $&   \citet{Melandri05} & \\
J & 0.46885 &$ 20.38 \pm 0.28          $&   This paper & CTIO 1.3m\\
J & 0.59837 &$ 20.59 \pm 0.19          $&   This paper & PAIRITEL\\
\hline
H & 0.59837 &$ 19.58 \pm 0.153        $&   This paper & PAIRITEL\\
\hline
K & 0.57962 &$ 18.53 \pm 0.176         $&   This paper & PAIRITEL \\
\enddata
\tablecomments{All optical and NIR observations used in fitting.  Data
were taken from the GCN circulars and our own observations using
Magellan, Keck (LRIS), and PAIRITEL.}
\tablenotetext{a}{Upper limit.}
\tablenotetext{b}{The NIC-FPS \citep{Vincent03} $Z$ was assumed to be
equivalent to the SDSS $z'$.}
\end{deluxetable}

\begin{deluxetable}{rrrrrrr}
\tablewidth{0pc}
\tablecaption{{\it Swift} XRT Aperture Photometry\label{t:xrtphot}}
\tablehead{
\colhead{Obs.} &
\colhead{Duration} &
\colhead{Exposure} &
\colhead{$C_\mathrm{A}$} &
\colhead{$C_\mathrm{B}$} &
\colhead{Bkg.\ count density} &
\colhead{GRB count rate}\\ &
\colhead{(s)} &
\colhead{(s)} &
&
&
\colhead{($10^{-2}$ counts pixel$^{-1}$)} &
\colhead{($10^{-3}$ counts s$^{-1}$)}}
\startdata
1a &     358.5 &    358.5 &  100 &    2 &   0.056 & 396.0 $\pm$ 39.7 \\
1b &    2221.1 &   2153.8 &  315 &   20 &   0.182 & 207.4 $\pm$ 11.7 \\
1c &    1385.7 &   1358.9 &  130 &    8 &   0.084 & 135.6 $\pm$ 11.9 \\
1d &    1031.2 &   1025.5 &   63 &    6 &   0.028 &  87.0 $\pm$ 11.0 \\
1e &    4441.6 &   1049.7 &   57 &    7 &   0.014 &  76.9 $\pm$ 10.2 \\
1f &     614.7 &    536.5 &   16 &    1 &   0.070 &  42.0 $\pm$ 10.6 \\
1g &    2396.5 &    409.4 &   14 &    1 &   0.042 &  48.3 $\pm$ 13.0 \\
1h &    2391.1 &    287.3 &    8 &    1 &   0.028 &  39.7 $\pm$ 14.0 \\
1i &    1176.2 &    332.8 &    6 &    1 &   0.000 &  25.5 $\pm$ 10.5 \\
1j &    1553.7 &    429.5 &   16 &    1 &   0.070 &  52.4 $\pm$ 13.3 \\
1k &    2394.0 &    659.2 &   13 &    1 &   0.014 &  27.9 $\pm$  7.8 \\
1l &    2388.4 &    639.3 &   22 &    5 &   0.056 &  48.3 $\pm$ 10.5 \\
1m &    1742.2 &    437.0 &    5 &    2 &   0.126 &  15.2 $\pm$  7.3 \\
2  &   52119.8 &   3399.1 &   19 &    2 &   0.294 &   7.68 $\pm$  1.83 \\
3  &   80975.1 &   2828.2 &   13 &    6 &   0.238 &   6.19 $\pm$  1.82 \\
4  &   59610.9 &   8361.7 &   25 &   10 &   0.630 &   3.98 $\pm$  0.85 \\
5  &   29603.5 &   3116.6 &    5 &    5 &   0.182 &   2.03 $\pm$  1.02 \\
6  &  180508.6 &  42828.5 &   27 &   80 &   2.18 &   0.65 $\pm$  0.17 \\
7  &   82821.9 &  21278.1 &   15 &   48 &   1.15 &   0.72 $\pm$  0.26 \\
8  &   86092.9 &  32161.4 &   24 &   51 &   2.14 &   0.79 $\pm$  0.22 \\
9  &   87416.5 &  33091.2 &   22 &   59 &   2.45 &   0.64 $\pm$  0.20 \\
10 &   35176.6 &   2943.5 &    0 &   12 &   0.22 &  --- \\
11 &  133401.7 &  40145.1 &   28 &  120 &   4.45 &   0.51 $\pm$  0.19 \\
12 &  167985.7 &  21764.1 &   16 &   50 &   4.09 &   0.40 $\pm$  0.26 \\
\enddata
\tablecomments{For each (sub)observation, we list the duration of the
XRT observation, the amount of exposure in the Photon Counting mode,
the counts in regions A and B (see Fig~\ref{f:xrtimg}), the space
density of background counts, and the estimate of the GRB count rate
using the expression in \S~\ref{s:xrtreduc}. The count rate
uncertainties are 1-$\sigma$.}
\end{deluxetable}

\begin{deluxetable}{rl}
\tablewidth{0pc}
\tablecaption{Broken Powerlaw Model Parametersfor the XRT\label{t:xdecline_fit}}
\tablehead{\colhead{Parameter} & \colhead{Value}}
\startdata
$\alpha_\mathrm{x,1}$ & 0.63\err{0.15}{0.19} \\
$\alpha_\mathrm{x,2}$ & 1.08\err{0.05}{0.04} \\
$t_\mathrm{break}$ & $1.26\err{1.19}{0.36}\times10^4$~s \\
$A$ & 59.4\err{160}{48.1}\\
\enddata
\tablecomments{90\% confidence intervals.}
\end{deluxetable}

 

\begin{deluxetable}{rrr}
\tablewidth{0pc}
\tablecaption{X-ray Spectral Fit Parameters\label{t:xrtspec}}
\tablehead{\colhead{Parameter} & \colhead{Before $t_\mathrm{break}$} & \colhead{After $t_\mathrm{break}$}}
\startdata
Photon Index $\Gamma$ & $2.31 \pm 0.75$ & $1.33 \pm 0.52$\\
$N_\mathrm{H}/10^{22}$~cm$^{-2}$ & 1.5\err{1.1}{0.9} & 0.52\err{0.45}{0.20}\\
\enddata
\tablecomments{90\% confidence intervals.}
\end{deluxetable}



\begin{deluxetable}{lccccccc}
\tablewidth{0pc}
\tablecaption{Absorption Line Summary \label{t:ewspec}}
\tabletypesize{\footnotesize}
\tablehead{\colhead{Ion} &\colhead{$\lambda_{\rm rest}$} & \colhead{W$_{\rm rest}$}  \\ &(\AA) & (m\AA) }
\startdata
\ion{Mg}{1} & 2852.964 & $1350 \pm 70$  \\
\ion{Ti}{2} & 3073.877 & $170 \pm 70$  \\
\ion{Ti}{2} & 3230.131 & $200 \pm 40$  \\
\ion{Ti}{2} & 3242.929 & $570 \pm 80$  \\
\ion{Ti}{2} & 3384.740 & $555 \pm 40$  \\
\ion{Fe}{1} & 3021.519 & $300 \pm 50$  \\
\enddata
\tablecomments{Errors in $W_r$ do not include uncertainty due to 
continuum placement.  For the weakest transitions, the systematic
error will be comparable to the statistical error. }
\end{deluxetable}

\begin{deluxetable}{llcll|lll}
\tablewidth{0pc}
\tablecaption{Closure Parameters and $\chi^2$ for Different Afterglow Models}
\tablehead{
 \colhead{}      & \colhead{}        & \colhead{}      & \colhead{}        &
\colhead{}    &\multicolumn{3}{c}{Fit with extinction} \\
 \colhead{Model} & \colhead{$\nu_c$} & \colhead{[b,c]} & \colhead{Closure} &
\colhead{$p$} & \colhead{$\beta$} & \colhead{$A_V$} & \colhead{$\chi^2 /
\mathrm{dof}$}}
\startdata
ISM  & B & [-3/2, 0]    & -1.16 $\pm$ 0.15 & 2.05 $\pm$ 0.04 & 0.525 & 0.567 &
7.82 / 5 \\
     & R & [-3/2, 1/2]  & -0.66 $\pm$ 0.15 & 1.71 $\pm$ 0.04 & 0.859 & 0.341 &
5.31 / 5 \\
Wind & B & [-3/2, -1/2] & -1.66 $\pm$ 0.15 & 1.38 $\pm$ 0.04 & 0.192 & 0.786 &
11.3 / 5 \\
     & R & [-3/2, 1/2]  & -0.66 $\pm$ 0.15 & 1.72 $\pm$ 0.04 & 0.859 & 0.341 &
5.31 / 5 \\
Jet  & B & [-2, -1]     & -2.81 $\pm$ 0.20 & 0.79 $\pm$ 0.03 & -0.10 & 0.972 &
15.3 / 5 \\
     & R & [-2, 0]      & -1.81 $\pm$ 0.20 & 0.79 $\pm$ 0.03 & 0.394 & 0.660 &
9.05 / 5 \\
\enddata
\tablecomments{Closure relations and parameters for a variety of
afterglow models, after \citet{Price02}.  None of the models are
well-supported by our data without corrections for host extinction.
The best-fit $\chi^2$ when fitting for host extinction is given in the
rightmost column; much better agreement is achieved.}
\end{deluxetable}

\clearpage

\end{document}